\documentclass[a4paper,fleqn]{cas-dc}
\usepackage[numbers]{natbib}
\usepackage[ruled,vlined,linesnumbered]{algorithm2e}
\usepackage{booktabs}
\usepackage{xcolor}
\usepackage{array}
\usepackage{adjustbox}
\usepackage{stfloats}
\usepackage{amssymb}
\usepackage{amsmath}
\usepackage{multirow,array}
\usepackage{graphicx}
\usepackage{ragged2e}
\usepackage{enumitem}
\usepackage{caption}
\usepackage{multirow}
\usepackage{rotating}
\usepackage{makecell}
\usepackage{tabularx}
\usepackage{lmodern}
\usepackage{rotating}
\usepackage{threeparttable}
\usepackage{longtable}
\usepackage{lscape} 
\usepackage{subcaption}
\usepackage{ltxtable}
\usepackage{tcolorbox}
\usepackage{tikz}
\usepackage{float}
\usepackage{listings}
\usepackage{hyperref}


\def\tsc#1{\csdef{#1}{\textsc{\lowercase{#1}}\xspace}}
\tsc{WGM}
\tsc{QE}
\begin{document}
\let\WriteBookmarks\relax
\def\floatpagepagefraction{1}
\def\textpagefraction{.001}
\let\printorcid\relax % 可去掉页面下方的ORCID(s)

% \shorttitle{<short title of the paper for running head>} 
\shorttitle{}    

% \shortauthors{<short author list for running head>}
\shortauthors{Zanxiang He et al.}

%论文标题
\title[mode = title]{\textsc{Mapis}: A Knowledge-Graph Grounded Multi-Agent Framework for Evidence-Based PCOS Diagnosis}  

%作者信息

\author[university1,university2]{Zanxiang He\fnref{equal}}
%\ead{hezanxiang2023@email.szu.edu.cn}
\author[university1]{Meng Li\fnref{equal}}
%\ead{limeng2@sztu.edu.cn}
\author[institute]{Liyun Shi\fnref{equal}}
%\ead{shi.liyun@szhospital.com}
\fntext[equal]{The three authors contribute equally to this work.}

\author[university1]{Weiye Dai}

\author[university1]{Liming Nie\corref{corresponding}}
\ead{nieliming@sztu.edu.cn}
\cortext[corresponding]{Corresponding author.}

\affiliation[university1]{
organization={Shenzhen Technology University},
%addressline={3002 Lantian Road, Pingshan District},
city={Shenzhen},
%postcode={518118},
state={Ghuangdong},
country={China}
}
\affiliation[university2]{
organization={Shenzhen University},
%addressline={3002 Lantian Road, Pingshan District},
city={Shenzhen},
%postcode={518118},
state={Ghuangdong},
country={China}
}
\affiliation[institute]{
organization={Shenzhen People's Hospital},
%addressline={1017 Dongmen North Road, Luohu District},
city={Shenzhen},
%postcode={518020},
state={Ghuangdong},
country={China}
}
\affiliation[university3]{
organization={University of South Australia},
%addressline={3002 Lantian Road, Pingshan District},
city={Adelaide},
%postcode={518118},
state={South Australia},
country={Australia}
}
\affiliation[university4]{
organization={Mahidol University},
%addressline={3002 Lantian Road, Pingshan District},
city={Krung Thep Maha Nakhon},
%postcode={518118},
state={Phitsanulok},
country={Thailand}
}
\affiliation[university5]{
organization={Sun Yat-sen University},
%addressline={135 Xingang West Road},
city={Guangzhou},
%postcode={510275},
state={Ghuangdong},
country={China}
}

\begin{abstract}
Polycystic Ovary Syndrome (PCOS) constitutes a significant public health issue affecting 10\% of reproductive-aged women, highlighting the critical importance of developing effective diagnostic tools. Previous machine learning and deep learning detection tools are constrained by their reliance on large-scale labeled data and an lack of interpretability. Although multi-agent systems have demonstrated robust capabilities, the potential of such systems for PCOS detection remains largely unexplored. Existing medical multi-agent frameworks are predominantly designed for general medical tasks, suffering from insufficient domain integration and a lack of specific domain knowledge. To address these challenges, we propose \textsc{Mapis}, the first knowledge-grounded multi-agent framework explicitly designed for guideline-based PCOS diagnosis. Specifically, it built upon the 2023 International Guideline into a structured collaborative workflow that simulates the clinical diagnostic process. It decouples complex diagnostic tasks across specialized agents: a gynecological endocrine agent and a radiology agent collaborative to verify inclusion criteria, while an exclusion agent strictly rules out other causes. Furthermore, we construct a comprehensive PCOS knowledge graph to ensure verifiable, evidence-based decision-making. Extensive experiments on public benchmarks and specialized clinical datasets, benchmarking against nine diverse baselines, demonstrate that \textsc{Mapis} significantly outperforms competitive methods. On the clinical dataset, it surpasses traditional machine learning models by 13.56\%, single-agent LLMs by 6.55\%, and previous medical multi-agent systems by 7.05\% in Accuracy.
\end{abstract}


% \begin{highlights}
% \item highlight-1
% \item highlight-2
% \item highlight-3
% \end{highlights}

% Keywords

\maketitle

% Main text
\section{Introduction}
Polycystic ovary syndrome (PCOS) is a significant public health issue characterized by endocrine, reproductive, cardiometabolic, dermatologic, and psychological manifestations \cite{azziz2016polycystic}. Despite affecting around 10\% of reproductive-aged women \cite{teede2018recommendations}, a substantial proportion remains undiagnosed \cite{march2010prevalence}. The clinical presentation of PCOS is highly heterogeneous. It encompasses psychological symptoms such as anxiety and depression, dermatologic signs including hirsutism and acne, reproductive dysfunctions ranging from irregular menstrual cycles to infertility, and metabolic comorbidities such as insulin resistance and type 2 diabetes \cite{teede2010polycystic,boomsma2006meta,apridonidze2005prevalence}. Clinical diagnosis currently follows the Rotterdam criteria, which require a structured synthesis of evidence to identify at least two of three cardinal features, clinical or biochemical hyperandrogenism, irregular menstrual cycles, and polycystic ovarian morphology (PCOM), while strictly excluding other causes \cite{eshre2004revised}.

However, despite these established standards, the diagnosis and management of PCOS remain clinically intricate. Difficulties in operationalizing individual diagnostic criteria, marked clinical heterogeneity, the influence of excess weight, ethnic differences, and variation across the life course all contribute to this complexity \cite{saini2016gaps}. These factors contribute to substantial variation in diagnosis, clinical presentation, and care pathways, which in turn leads to delayed diagnosis, a poor diagnostic experience, and widespread dissatisfaction with care, as reported by women internationally \cite{gibson2017delayed}. Consequently, the high cognitive load and inherent heterogeneity in current practice underscore an urgent need for advanced diagnostic tools that assist clinicians in delivering precise, consistent, and guideline-adherent care.

Previous studies in detecting PCOS have primarily utilized deep learning and machine learning approaches \cite{aggarwal2023early,tiwari2022sposds,65chauhan2021comparative,46abouhawwash2023automatic}. However, these approaches are inherently constrained by their dependence on large volumes of labeled data, limited interpretability, and a primary focus on statistical patterns rather than clinical causal reasoning \cite{li2025intelligent}. Such opacity fails to meet the stringent requirements for evidence-based decision-making in medical practice and renders them ill-suited for processing unstructured clinical narratives without extensive manual feature engineering. Recently, Large Language Models (LLMs) and several multi-agent frameworks have emerged in the broader medical domain, demonstrating improved collaborative reasoning in general diagnostic tasks \cite{kim2024mdagents,tang-etal-2024-medagents,li2024agent}. However, despite these advancements, their specific potential for PCOS detection remains largely unexplored. Current systems remain generic and lack the specialization required for guideline-driven disorders like PCOS, where precise adherence to criteria such as the Rotterdam consensus is critical. Moreover, deploying LLMs for such tasks faces the significant challenge of model hallucination, a risk exacerbated by the current absence of a PCOS-specific Knowledge Graph to ground generative reasoning in authoritative facts.

To address these challenges, we propose \underline{\textbf{M}}ulti-\underline{\textbf{A}}gent \underline{\textbf{P}}COS \underline{\textbf{I}}ntelligent Detection \underline{\textbf{S}}ystem (\textsc{Mapis}), a novel framework built upon International Evidence-based Guideline for the assessment and management of polycystic ovary syndrome 2023 \cite{teede2023international} that explicitly mirrors the clinical diagnostic workflow through multi-agent collaboration grounded in domain knowledge graph. Specifically, \textsc{Mapis} integrates five synergistic modules. The framework initiates with the data preprocessing module converting raw EHRs into structured formats, grounded by a knowledge graph construction module that mitigates hallucination risks by encoding authoritative guidelines. Central to the architecture, the three-step assessment and exclusion modules execute the diagnostic logic through specialized collaboration: a coordinator agent orchestrates the gynecological endocrine agent to verify irregular cycles and hyperandrogenism (Steps 1-2), and the radiology agent to evaluate polycystic ovarian morphology (Step 3). Subsequently, an exclusion agent enforces the mandatory exclusion phase to strictly rule out other causes before validation. Finally, the report generation module synthesizes these findings into a transparent clinical report, enabling reliable, zero-shot diagnostic inference independent of large-scale training data. 

To validate the efficacy of \textsc{Mapis}, we conducted extensive evaluations on both public benchmarks and a private clinical dataset, benchmarking against nine diverse baselines. Experimental results demonstrate that \textsc{Mapis} significantly outperforming competitive methods. Specifically, on the clinical dataset, it surpasses traditional machine learning models by 13.56\%, single-agent by 6.55\%, and previous medical multi-agent systems by 7.05\% in Accuracy. We present a scalable, transferable paradigm: this framework can be generalized to other guideline-dependent disorders by constructing specialized multi-agent systems that built upon the specific diagnostic workflows mandated by their respective guidelines, thereby ensuring effective and standardized diagnosis.

%Our approach directly mirrors the rigorous workflow of clinical diagnosis: a Coordinator Agent orchestrates the systematic three-step Rotterdam evaluation sequence and dynamically distributing patient data to specialized agents. Specifically, a Gynecological Endocrine Agent evaluates irregular menstrual cycles and clinical hyperandrogenism (Step 1), and proceeding to biochemical assessment only when clinical signs are insufficient (Step 2); a Radiology Agent interprets ultrasound findings for polycystic ovarian morphology (Step 3); and finally, an Exclusion Agent conducts a mandatory differential exclusion phase to ensure diagnostic validity. Crucially, each agent is augmented by a comprehensive PCOS-specific knowledge graph (KG) constructed from international guidelines and expert consensus. By querying this graph through hierarchical entity linking and ``U-Retrieval'', the agents access precise diagnostic thresholds, age-specific criteria, and exclusion requirements. This architecture enables zero-shot diagnostic inference without labeled training data from target populations. The agents reason over guideline-encoded medical knowledge rather than learned statistical patterns, which ensures consistent, standard-compliant assessment across varied patient profiles, independent of the statistical biases inherent in localized training data. 

The main contributions are as follows:
\begin{itemize}
    \item [\textbullet] To the best of our knowledge, we present the first guideline-based multi-agent framework explicitly designed for PCOS diagnosis. By strictly operationalizing clinical protocols into a collaborative agent workflow, our system effectively addresses the dual challenges of diagnostic complexity and data scarcity.
    \item [\textbullet] We construct the first domain knowledge graph for PCOS. Synthesized from authoritative guidelines, this structured knowledge base serves as a determinisitc external memory, mitigating model hallucination by grounding agent responses in precise, evidence-based medical facts.
    \item [\textbullet] Extensive experiments on both public and private clinical datasets demonstrate superior performance. \textsc{Mapis} significantly outperforms baselines, including traditional machine learning models, single-agent, and existing multi-agent systems, thereby demonstrating the efficacy of our guideline-based clinical reasoning framework.

\end{itemize}

\section{Related Work}
\subsection{Intelligent PCOS Detection}
PCOS has become a focal point of medical AI research, driven by the urgent need for early, accurate screening tools. Early Machine Learning classifiers, including SVMs \cite{72deshpande2014automated}, Decision Trees \cite{65chauhan2021comparative}, and Ensemble methods \cite{67zhang2021raman}, established baselines but struggled with manual feature engineering and non-linear interactions. 

Consequently, recent research has increasingly shifted towards deep learning paradigms to enhance diagnostic performance. One dominant stream utilizes Convolutional Neural Networks (CNNs) for automated feature extraction from ultrasound \cite{srivastav2024transfer} or scleral images \cite{56lv2022deep}, while another stream employs architectures like Bi-LSTM and MLP \cite{46abouhawwash2023automatic} to process complex structured clinical data. To further optimize these models, auxiliary strategies such as Principal Component Analysis (PCA) \cite{55zigarelli2022machine} and Swarm Intelligence algorithms \cite{18subha2024computational} have been integrated to reduce dimensionality and identify significant biomarkers. 

However, despite these advancements, these methods are limited by their reliance on large-scale labeled data and inherent lack of interpretability. They prioritize statistical correlation over clinical reasoning, failing to provide the verifiable evidence chains. Furthermore, without complex manual feature engineering, they struggle to process unstructured clinical narratives, such as detailed patient histories. In this work, we introduce a guideline-grounded multi-agent framework that simulates the clinical diagnostic process to achieve interpretable, zero-shot diagnosis without reliance on large-scale training data.

\begin{figure*}[H] % 这里的 * 号是关键
    \centering
    % width=\textwidth 确保图片宽度占满整个页面宽度（两栏总宽）
    \includegraphics[width=\textwidth]{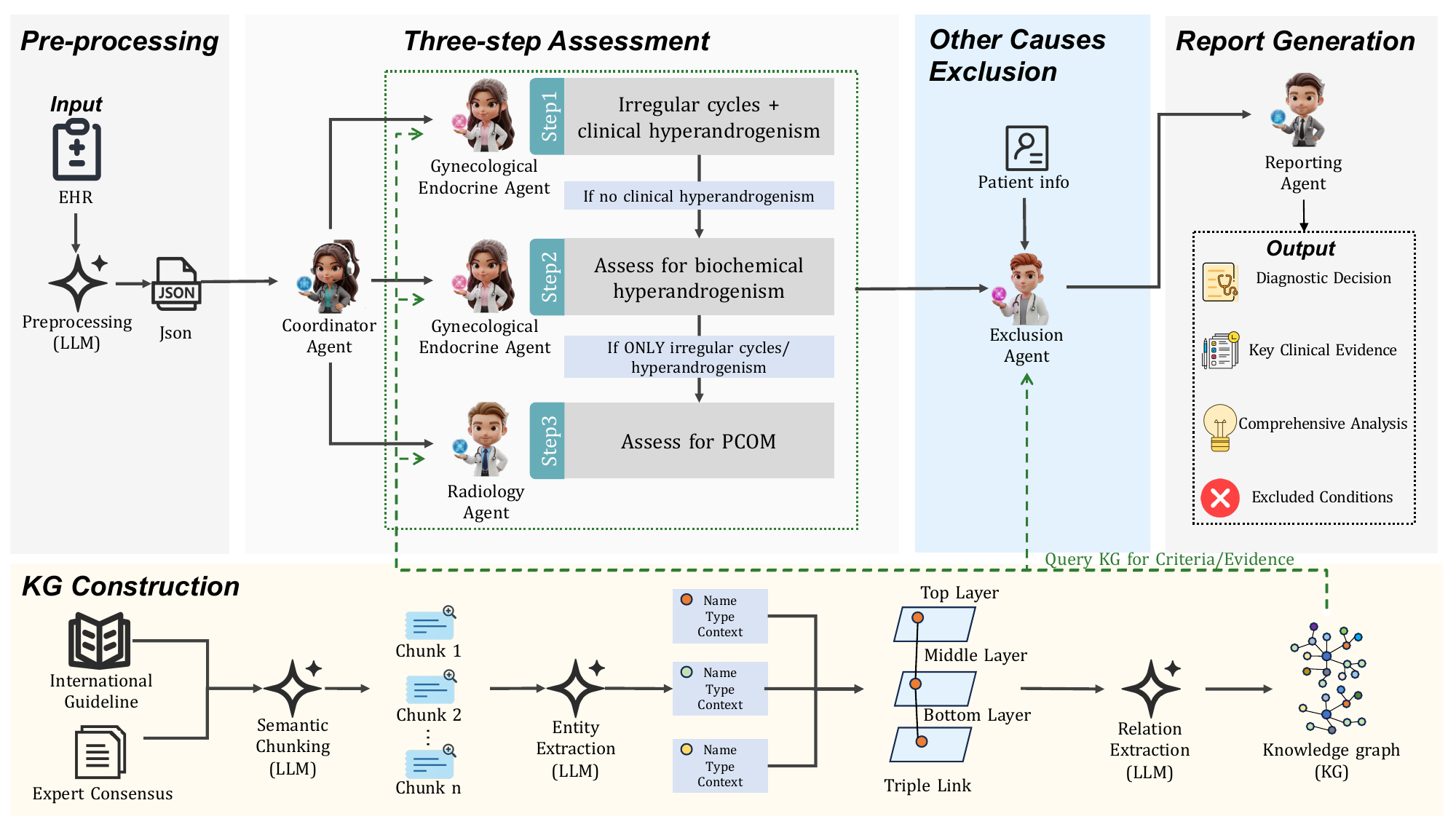}
    \caption{Schematic overview of \textsc{Mapis}. The architecture is organized into two core processes integrating five synergistic modules: (1) Knowledge Construction, where the KG Construction Module transforms guidelines into a structured knowledge base; and (2) Diagnostic Execution, which proceeds through the Clinical Data Preprocessing, Three-Step Assessment, Exclusion Modules, and Report Generation Module.}
    \label{fig1}
\end{figure*}
\subsection{Multi-Agent Systems in Medical Decision Making}
The emergence of large language model-based multi-agent systems has revolutionized medical diagnosis by enabling collaborative reasoning through division of labor and specialized expertise coordination. Multi-agent architectures offer unique advantages in handling complex medical decision-making: they improve efficiency through task decomposition across agents responsible for information retrieval, knowledge extraction, and result integration; provide scalability to adapt to evolving medical knowledge; and enhance interpretability by making each processing step traceable through clearly defined agent roles.

Recent advances have introduced increasingly sophisticated collaborative frameworks. MedAgents \cite{tang-etal-2024-medagents} employs role-playing agents in multi-round discussions to reach diagnostic consensus through iterative analysis, while MDAgents \cite{kim2024mdagents} 
advances this paradigm by automatically assigning collaboration structures—switching between solo and group analysis—tailored to task complexity. Pushing the boundaries of simulation, ``Agent Hospital'' \cite{li2024agent} a simulacrum where autonomous doctor agents evolve by treating simulated patients, achieving state-of-the-art performance through self-evolution without manual labeling. Similarly, mimicking real-world clinical referral systems, Agent-derived Multi-Specialist Consultation (AMSC) framework \cite{wang2024beyond}, which models the diagnosis process by consulting specialists and adaptively fusing their probability distributions over potential diseases.

Despite these promising developments, the potential of collaborative AI architectures for PCOS detection remains largely unexplored. Existing multi-agent medical systems function primarily as general-purpose frameworks, suffering from insufficient domain integration. They predominantly rely on generic heuristic coordination mechanisms (such as majority voting, open-ended evolution, or probabilistic fusion) rather than explicitly operationalizing the disease-specific diagnostic workflows mandated by clinical guidelines. This limitation is particularly acute for conditions like PCOS, which require a systematic, multi-step protocol. In this work, we propose a novel framework that leverages domain knowledge grounding, guideline-based workflow operationalization, and systematic differential exclusion to enable reliable PCOS detection through collaborative reasoning among specialized agents.

\subsection{Knowledge-Enhanced Medical Reasoning}
Retrieval-Augmented Generation (RAG) has become a standard paradigm for mitigating hallucinations in LLMs by grounding generation in external data without retraining \cite{lewis2020retrieval,guu2020retrieval}. However, in clinical settings, standard RAG often falls short as it retrieves unstructured text chunks that lack the evidence-based reasoning and precise term explanations required for medical decision-making \cite{miao2024integrating}.

To overcome the limitations of flat text retrieval, recent advances have integrated knowledge graph to provide structured, inferable information \cite{8530028,izacard2021unsupervised}. GraphRAG \cite{edge2024local} represents a significant leap forward , enhancing complex reasoning by organizing retrieval data into graph structures rather than isolated snippets. Adapting the graph paradigm specifically for healthcare, MedGraphRAG \cite{wu-etal-2025-medical} effectively improves the reliability of question answering in the general medical field. It introduces a ``Triple Graph Construction'' method that integrates user private data, authoritative medical literature, and UMLS medical dictionaries to build a hierarchical triple knowledge graph. Furthermore, it employs a ``U-Retrieval'' strategy to efficiently balance global context awareness with precise entity indexing.

However, there is currently a lack of domain-specific knowledge graphs in the specialized vertical field of PCOS diagnosis. Given the inherent hallucination issues of LLMs and the extremely high requirements for Accuracy in clinical diagnosis using RAG, we constructed a PCOS-specific knowledge graph based on the MedGraphRAG methodology. This initiative aims to address the Accuracy and traceability of model-generated content by introducing external evidence-based knowledge.

%While MedGraphRAG excels at generating evidence-based responses for general medical queries, it functions primarily as an evidence retriever rather than a diagnostic inference engine. It lacks the capability to handle the conditional evaluation logic and mandatory exclusion pathways required for syndrome-specific diagnosis. Diagnosing complex conditions like PCOS demands a ``Diagnostic Rule Base'' capable of processing age-specific thresholds and systematic differential exclusion. Building upon the hierarchical efficiency of MedGraphRAG, our work adapts its three-tier architecture specifically for Guideline Operationalization: rather than linking general documents, our PCOS-centric knowledge graph structurally encodes the diagnostic dependency chain, integrating expert heuristics and the 2023 International Guideline to enable precise, zero-shot clinical reasoning that generalist graph systems cannot achieve.

\section{Method}

\subsection{Overview}
Inspired by the clinical decision pathway outlined in the 2023 International Guideline for the Assessment and Management of Polycystic Ovary Syndrome \cite{teede2023international}, we propose \textsc{Mapis}, a multi-agent framework explicitly designed to simulate the clinical reasoning and diagnostic process of specialists. \textsc{Mapis} operationalizes the guideline-mandated diagnostic algorithm as a collaborative workflow to ensure precision and reliability. As illustrated in Figure \ref{fig1}, the framework comprises five modules:
\begin{itemize}
    \item [\textbullet] \textbf{Preprocessing:} Because clinicians must synthesize information from diverse sources, the system accepts raw EHRs as input. A preprocessing module using LLMs extracts and standardizes this heterogeneous data into a structured JSON format, ensuring consistent clinical representation across different documentation styles.
    
    \item [\textbullet] \textbf{Domain knowledge graph construction:} To mitigate the hallucination risks inherent in LLMs, we construct a comprehensive PCOS-specific knowledge graph. As illustrated in the KG Construction panel of Figure \ref{fig1}, this pipeline processes authoritative guidelines and papers through semantic chunking, followed by LLM-based entity-relation extraction and subgraph merging. This structured KG serves as an external, authoritative knowledge base that grounds the agents’ reasoning with precise clinical evidence.
 
    \item [\textbullet] \textbf{Three-step diagnostic assessment:} This core module implements the sequential evaluation framework. A Coordinator Agent orchestrates the workflow by dispatching tasks to specialized agents: Step 1: A Gynecological Endocrine Agent assesses irregular cycles and clinical hyperandrogenism; Step 2: If clinical evidence is insufficient, the agent proceeds to verify biochemical hyperandrogenism; Step 3: If only hyperandrogenism or only irregular cycles is present, a Radiology Agent evaluates ultrasound findings for polycystic ovarian morphology.
    
    \item [\textbullet] \textbf{Other causes exclusion:} Reflecting the critical mandate to ``exclude other causes,'' the workflow incorporates a dedicated exclusion gate. An Exclusion Agent conducts a differential diagnosis to rigorously rule out mimicking conditions, such as congenital adrenal hyperplasia (CAH), thyroid disorders, and hyperprolactinemia, before any positive diagnosis is confirmed.

    \item [\textbullet] \textbf{Report generation:} Finally, a reporting agent synthesizes the multistep findings to generate a comprehensive clinical report. This output includes detailed evidence for each criterion, outcomes of the exclusion analysis, and lifestyle management recommendations, supporting transparent and evidence-based decision-making.
\end{itemize}
    
\subsection{Multi-Agent}
Our system comprises five specialized agents, each engineered to execute a distinct component of the diagnostic pipeline, ranging from domain-specific clinical reasoning and differential exclusion to system orchestration and reporting. The collaborative interaction among these agents follows the workflow detailed in Section \ref{workflow}.

\subsubsection{Multi-Agent Workflow}\label{workflow}
The workflow operationalizes the sequential diagnostic pathway mandated by the 2023 International Guideline, orchestrated centrally by the coordinator agent. The process is structured into two logical phases: Criteria Verification and Differential Exclusion. Initially, the gynecological endocrine and radiology agents evaluate the Rotterdam inclusion criteria. Upon satisfying the definitive threshold (at least two of three criteria), the workflow transitions to a mandatory exclusion phase, where the exclusion agent rigorously validates the diagnosis against other causes. To optimize computational efficiency, the coordinator implements an adaptive execution strategy (Early Termination): resource-intensive steps, such as radiology analysis or full exclusion screening, are triggered conditionally—only if the preceding evaluations maintain the probabilistic viability of a positive diagnosis. Conversely, if definitive negative evidence is established early (rendering the ``two-out-of-three'' threshold mathematically unreachable), the process terminates immediately to prevent redundant computation. The formal control logic is detailed in Algorithm \ref{alg:workflow}.

\begin{algorithm}[htbp!]
\caption{Multi-Agent PCOS Diagnostic Workflow}
\label{alg:workflow}
\KwIn{$\mathcal{D}$ (Patient EHR Data), $\mathcal{G}$ (Knowledge Graph)}
\KwOut{$R$ (Clinical Report)}
\textbf{Initialize} Agents: $A_{\text{Coord}}, A_{\text{Endo}}, A_{\text{Rad}}, A_{\text{Exc}}, A_{\text{Report}}$\;
$\textit{Candidate} \leftarrow \text{False}$\tcp{Initialize diagnosis candidacy}
\BlankLine
\textbf{Phase 1: Sequential Rotterdam Assessment}\;
\tcc{Step 1: Assess Irregular Cycles and Clinical Hyperandrogenism}
$C_{\text{cyc}}, C_{\text{clin}} \leftarrow A_{\text{Endo}}.\text{evaluate}(\mathcal{D}, \mathcal{G})$\;
\eIf{$C_{\text{cyc}} \land C_{\text{clin}}$}{
    $\textit{Candidate} \leftarrow \text{True}$\;
}{
    \tcc{Step 2: Assess Biochemical Hyperandrogenism}
    $C_{\text{bio}} \leftarrow A_{\text{Endo}}.\text{evaluate\_bio}(\mathcal{D}, \mathcal{G})$\;
    \eIf{$C_{\text{cyc}} \land C_{\text{bio}}$}{
        $\textit{Candidate} \leftarrow \text{True}$\;
    }{
        \tcc{Step 3: Assess PCOM only if needed}
        $C_{\text{pcom}} \leftarrow A_{\text{Rad}}.\text{evaluate\_imaging}(\mathcal{D}, \mathcal{G})$\;
        \If{$(C_{\text{cyc}} \lor C_{\text{clin}} \lor C_{\text{bio}}) \land C_{\text{pcom}}$}{
            $\textit{Candidate} \leftarrow \text{True}$\;
        }
    }
}
\BlankLine
\textbf{Phase 2: Exclusion and Reporting}\;
\eIf{$\textit{Candidate}$ is True}{
    \tcc{Mandatory Exclusion of differentials (e.g., CAH, Cushing's)}
    $\textit{ExclusionResult} \leftarrow A_{\text{Exc}}.\text{exclude\_differentials}(\mathcal{D}, \mathcal{G})$\;
    \eIf{$\textit{ExclusionResult}.\text{passed}$}{
        $\textit{FinalDiagnosis} \leftarrow \text{PCOS\_CONFIRMED}$\;
    }{
        $\textit{FinalDiagnosis} \leftarrow \text{ALTERNATIVE: } + \textit{ExclusionResult}.\text{cause}$\;
    }
}{
    $\textit{FinalDiagnosis} \leftarrow \text{PCOS\_EXCLUDED}$\;
}
$R \leftarrow A_{\text{Report}}.\text{generate}(\text{AllFindings}, \textit{FinalDiagnosis})$\;
\Return{$R$}
\end{algorithm}

\subsubsection{Agent Roles}
The responsibilities and capabilities of each agent are defined as follows:

\paragraph{Coordinator Agent} Acting as the central orchestrator, the coordinator agent is responsible for data distribution and workflow control. It parses structured patient data from EHRs and determines the information requirements for each diagnostic step. By maintaining the global session state, the coordinator manages the sequential three-step assessment and dynamically routes the workflow based on intermediate outcomes.

\paragraph{Gynecological Endocrine Agent} This agent specializes in the first two inclusion criteria of the Rotterdam consensus. In Step 1, it analyzes menstrual history against guideline-defined thresholds to identify irregular menstrual cycles. In Step 2, it evaluates hyperandrogenism through a hierarchical approach: initially assessing clinical phenotypes and, if clinical signs are absent or ambiguous, proceeding to verify biochemical markers to confirm androgen excess.

\paragraph{Radiology Agent} The radiology agent specializes in Step 3 of the diagnostic process. It analyzes descriptions of ultrasound findings to assess PCOM. The agent interprets imaging parameters according to thresholds defined in the 2023 guideline, including follicle count per ovary and ovarian volume, to determine whether the morphological criterion is met.

\paragraph{Exclusion Agent} Acting as a diagnostic safety gatekeeper, this agent executes the mandatory differential exclusion phase. It rigorously analyzes specific laboratory indicators to systematically rule out other causes—including Non-classic Congenital Adrenal Hyperplasia (NCCAH), thyroid dysfunction, and hyperprolactinemia—ensuring that the final diagnosis is specific and robust against confounders.

\paragraph{Reporting Agent} Serving as the final synthesizer, this agent aggregates findings from all specialized agents to generate a comprehensive diagnostic report. It applies the ``two-out-of-three'' rule to derive the final decision, explicitly documenting the status of each criterion and the results of the exclusion analysis. Additionally, it provides risk stratification and clinical recommendations for follow-up, facilitating actionable patient management.

\subsubsection{Agent Prompt Templates}
To ensure robust and clinically verifiable reasoning, we employ a modular ``Role-Guideline-Task-Constraint'' prompting paradigm. By replacing open-ended generation with a rigorous template architecture, we strictly bound the LLM's reasoning space within authoritative medical standards. As illustrated in Figure \ref{fig:prompt_example}, the prompt structure comprises four standardized components:
    \paragraph{System Role Definition} Establishes a specialized clinical persona (e.g., Senior Gynecological Endocrinologist) to activate domain-specific latent knowledge. This ensures the model adopts the appropriate professional stance and terminological precision required for medical inference.
    \paragraph{Contextual Knowledge Injection} Rather than relying on the model's parametric memory, this module dynamically injects precise diagnostic criteria retrieved from our PCOS KG. As depicted in the Diagnostic Guidelines section of the prompt, exact parameters—such as age-specific menstrual irregularity thresholds or Ferriman-Gallwey cutoffs—are explicitly embedded. This mechanism forces the agent to ground its evaluation in deterministic medical logic rather than probabilistic approximation.
    \paragraph{Task Specification} Delineates the atomic objective for the specific agent. This ranges from verifying the ``two-out-of-three'' Rotterdam inclusion criteria to executing a systematic differential exclusion protocol, ensuring each agent focuses on a single, manageable aspect of the diagnostic pipeline.
    \paragraph{Structured Output Constraint} Enforces a rigid, nested JSON output schema. This constraint compels the agent to perform explicit chain-of-thought reasoning—articulating the ``Status'' and ``Reasoning'' for each criterion—before aggregating the final conclusion. This structure not only enhances interpretability but also ensures the output is machine-parsable for the coordinator agent's downstream synthesis.

% Place this code at the very end of your .tex file, 
% typically after \bibliography{...} or before \end{document}

% If you want it in an appendix section, uncomment the next two lines:
% \appendix
% \section{Prompt Templates}

\begin{figure*}[t!] % Changed to figure* for double-column span
    \centering
    \begin{tcolorbox}[
        colback=gray!5,
        colframe=black!70,
        arc=2mm,
        boxrule=0.8pt,
        fontupper=\small,
        title=\textbf{Prompt for Gynecological Endocrine Agent (Step 1)},
        coltitle=white,
    ]
    
    \textbf{[SYSTEM ROLE]} \\
    You are a Senior Consultant Gynecological Endocrinologist. Your core function is to analyze structured patient data against Rotterdam criteria.

    \vspace{0.4em}
    \textbf{[TASK INPUT]} \\
    Your task is to assess 2 of the 3 Rotterdam criteria components (irregular cycles AND Clinical hyperandrogenism), based on guidelines, and clinical judgment from the patient data. This is an initial step in a two-of-three criteria evaluation, not the definitive diagnosis. \\
    Patient Data JSON: \texttt{\{...menstrual\_history, skin\_exam...\}}

    \vspace{0.4em}
    \textbf{[DIAGNOSTIC GUIDELINES (Static Rules)]}
    \begin{itemize}[leftmargin=1.2em, nosep]
        \item \textbf{Irregular Cycles:} $>3$ years post-menarche: Irregular if $<21$ or $>35$ days OR $<8$ cycles/year. Any single cycle $>90$ days is irregular.
        \item \textbf{Hyperandrogenism:} Primary: Hirsutism (Ferriman-Gallwey $\ge 2$). Secondary: Acne/Alopecia (weak when isolated).
    \end{itemize}

    \vspace{0.4em}
    \textbf{[KNOWLEDGE BASE INFORMATION (Dynamic Injection)]}
    \texttt{\{graphrag\_knowledge\}}
    \textcolor{gray}{\small $\leftarrow$ Retrieved from KG}

    \vspace{0.4em}
    \textbf{[OUTPUT CONSTRAINT]} \\
    Output a single JSON object evaluating the criteria:
    \begin{verbatim}
{
  1:Irregular_cycles: {
      status: Yes/No/Uncertain, 
      reasoning: A brief medical rationale strictly following
       the guidelines above.
  },
  2:Clinical_hyperandrogenism: {
      status: Yes/No/Uncertain, 
      reasoning: A brief medical rationale weighing the provi-
      ded signs according to the notes above.
  }
}
    \end{verbatim}
    \end{tcolorbox}
    \caption{An example of the prompt for the Gynecological Endocrine Agent (Step 1).}
    \label{fig:prompt_example}
\end{figure*}

\subsection{Domain Knowledge Graph}\label{knowledgegraph}
To mitigate the hallucination risks inherent in LLMs and ensure clinical precision, we construct a specialized hierarchical PCOS knowledge graph. Adopting the architecture of MedGraphRAG \cite{wu-etal-2025-medical}, our framework establishes a triple graph structure: a bottom layer anchored by our self-constructed PCOS Dictionary ($\mathcal{D}$); a middle layer derived from the 2023 International Guidelines and Expert Consensus; a top layer integrating private EHR data.

\subsubsection{PCOS Knowledge Graph Construction}
As illustrated in Figure \ref{fig1}, the construction pipeline follows a strictly sequential ``triple graph construction''~\cite{wu-etal-2025-medical} process to transform unstructured clinical texts into a computable logic network.

\paragraph{Semantic Chunking and Context-Aware Entity Extraction}
We process the authoritative source documents (Middle Layer) using a dynamic semantic chunking strategy. To preserve diagnostic logic flow, we utilize an embedding model to detect semantic discontinuities between paragraphs. A document is segmented into chunks $C = \{c_1, ..., c_n\}$ only when the topic context shifts. A buffer is applied to ensure that each chunk encapsulates a self-contained medical context, which is prerequisite for accurate dictionary alignment.

For each semantic chunk, we employ a Large Language Model ($L_E$) to extract entities. We prompt $L_E$ to valid entities and generate a structured output $e$ composed of three specific components:
\begin{equation}
e = \{name, type, context\}
\end{equation}
\begin{itemize}
    \item \textbf{Name ($na$)}: The precise textual mention from the document.
    \item \textbf{Context ($cx$)}: A descriptive summary generated by $L_E$ that contextualizes the entity within the chunk.
    \item \textbf{Type ($ty$)}: We constrain this slot using our specialized PCOS Ontology ($\Delta$) to ensure diagnostic rigor.  
\end{itemize}
    
\paragraph{Triple Layer Linking}
This phase implements the core hierarchical alignment mechanism. We replace the generic UMLS vocabulary used in MedGraphRAG with our specialized PCOS dictionary ($\mathcal{D}$) to enforce stricter domain constraints. Bottom-Middle Linking (Dictionary Grounding): Extracted entities from the guidelines (Middle Layer) are deterministically linked to the Bottom Layer ($\mathcal{D}$) via alias matching and semantic type constraints. For instance, varying terms like ``PCO'' or ``Polycystic Ovaries'' are normalized to the standard node defined in our dictionary, ensuring terminological consistency; Top-Middle Linking (EHR Integration): During the inference phase, extracted clinical entities from patient EHRs (Top Layer) are linked to the most relevant guideline nodes (Middle Layer) through embedding similarity. This establishes a connected pathway, enabling the system to trace patient symptoms back to authoritative rules and standard definitions.

\paragraph{Relation Extraction and Subgraph Generation}
In the final step, the model identifies directed diagnostic dependencies $(e_{head}, r, e_{tail})$ strictly within the boundaries of each semantic chunk. By conditioning the extraction on the localized \textit{context} generated in Step 1, the model captures precise logical rules rather than mere co-occurrences. These local subgraphs are then aggregated to form the comprehensive domain knowledge graph.

\subsubsection{Graph-Augmented Retrieval}
We employ a retrieval strategy inspired by the ``U-Retrieval'' mechanism \cite{wu-etal-2025-medical} to support agent reasoning. When an agent queries the system: Top-Down Indexing: The system first identifies the relevant high-level clinical concepts in the graph based on the query semantics; Contextual Retrieval: It then retrieves the specific diagnostic rules and thresholds stored in the \textit{context} fields of the linked nodes; Evidence Grounding: Finally, the retrieved knowledge is presented to the agent alongside its source citations from the guideline, ensuring that every diagnostic decision is verifiable and grounded in established medical evidence.

\section{Experimental Settings}
We present the experimental setup across four key dimensions: datasets, baselines, implementation details, and evaluation metrics.
\subsection{Datasets}
We evaluate the framework using two datasets: a widely used public benchmark to assess generalizability and a high-dimensional private clinical dataset to assess diagnostic precision in real-world settings.

\subsubsection{Public Dataset} We use the public PCOS dataset collected from 10 hospitals in Kerala, India \cite{pcosdataset}. This dataset serves as a benchmark in PCOS research \cite{18subha2024computational,34elmannai2023polycystic,57nasim2022novel,84kumari2023sms}, comprising 541 cases (364 PCOS-positive, 177 PCOS-negative) with 45 features covering demographics and basic physiological indicators. While valuable for baseline comparisons, a clinical review identified significant limitations: the dataset lacks critical biochemical markers mandated by international guidelines, specifically free testosterone and Dehydroepiandrosterone sulfate (DHEA-S), limiting its utility for validating hormonal exclusion logic.

\subsubsection{Private Dataset}
To address these limitations and support rigorous clinical validation, we curated the Gynecological Endocrine Dataset (GED) from Shenzhen People's Hospital (August–October 2025). This dataset comprises 170 distinct clinical cases. To simulate a realistic and challenging diagnostic scenario, the cohort includes 100 confirmed PCOS cases and a specific control group of 70 patients with Abnormal Uterine Bleeding-Ovulatory Dysfunction (AUB-O). The inclusion of AUB-O as the control group is strategic; it serves as a robust test for the system's differential diagnosis capabilities against etiologies that clinically mimic PCOS.

Distinguishing it from public datasets, GED encompasses 108 clinical features as defined in the literature \cite{li2025intelligent}. Crucially, it incorporates the mandatory biochemical indicators previously missing in public data, including free testosterone, DHEA-S, alongside comprehensive metabolic indices and standardized ultrasound parameters derived from raw EHRs. Data quality was strictly controlled by including only patients with complete, structured records at initial diagnosis. All data were anonymized in compliance with privacy standards, and the study protocol received approval from the Institutional Review Board.

\subsection{Baselines} 
To comprehensively evaluate the effectiveness of \textsc{Mapis}, we benchmark it against a total of nine diverse baselines spanning three distinct categories: traditional machine learning methods, single-agent-based approaches with varied prompting strategies, and existing medical multi-agent frameworks.

For traditional machine learning methods, we selected three representative algorithms with strong performance in prior PCOS literature: Bayesian classifiers \cite{ACritical}, which employ probabilistic graphical models to capture dependencies between clinical features; random forest with PCA-based feature selection (RF-PCA) \cite{i-HOPE}, which combines ensemble learning with dimensionality reduction for high-dimensional clinical data; and XGBoost-SCA \cite{Sinecosine}, which uses gradient boosting optimized with the sine–cosine algorithm, representing the current state of the art in traditional ML for PCOS detection.

To assess the necessity of multi-agent collaboration, we evaluated four advanced foundation models: GPT-4o and GPT-4.1 (OpenAI) \cite{hurst2024gpt}, alongside Gemini-2.5-Flash and Gemini-2.5-Pro (Google) \cite{comanici2025gemini}. Each model was evaluated using three inference strategies: zero-shot (direct diagnosis from patient data without external guidance or intermediate reasoning); chain-of-thought (CoT; prompting step-by-step reasoning before the conclusion); and standard RAG. Unlike our system's graph-based approach, this RAG baseline augments the single-agent setup by retrieving unstructured text chunks from the guideline documents via semantic similarity. This serves as a comparative baseline to determine whether simple access to unstructured knowledge is sufficient for diagnosis, isolating the specific contributions of our structured knowledge graph and multi-agent collaborative architecture. 

We compared \textsc{Mapis} against two general-purpose medical agent systems: MedAgent \cite{tang-etal-2024-medagents}, a multidisciplinary consultation framework that employs adaptive expert recruitment and majority-vote consensus; and MDAgents \cite{kim2024mdagents}, a collaborative system for differential diagnosis featuring hierarchical agent organization and progressive-refinement discussions.

\subsection{Implementation Details}
For traditional machine-learning baselines, we applied standardized preprocessing pipelines to both the Kerala and GED datasets. Missing values were imputed with the median for continuous variables and the mode for categorical variables. Continuous features were normalized to zero mean and unit variance using z-score standardization.
For the public Kerala dataset, we employed a stratified 80:20 train–test split. For our private GED dataset, given its smaller sample size ($N=170$), we utilized 5-fold cross-validation to ensure robust performance estimation.
Hyperparameter optimization used grid search to identify optimal configurations for each algorithm: Bayesian Network: Structure learning was performed using the Hill-Climbing algorithm with BIC scoring; Random forest with PCA: we tuned the number of estimators (100–500) and maximum depth, and selected PCA components to retain 95\% variance; XGBoost with SCA: We optimized the gradient-boosting hyperparameters (learning rate, maximum depth, subsample ratio) employing the sine–cosine algorithm as proposed in paper \cite{Sinecosine}.

For single-agent and multi-agent experiments, we employed four foundation models: GPT-4o (2024-11-20) and GPT-4.1 (2025-04-14) via the OpenAI API, alongside Gemini-2.5-Flash and Gemini-2.5-Pro via the Google AI Studio API. To ensure reproducibility and eliminate generative randomness, we set temperature to 0 and disabled top‑p sampling for all API calls.

Since existing medical multi-agent frameworks such as MedAgent \cite{tang-etal-2024-medagents} and MDAgents \cite{kim2024mdagents} were designed for open-ended medical question answering (QA) rather than structured classification, we adapted the inputs to their paradigms. Specifically, we serialized each patient’s structured feature vector into a natural-language clinical vignette. We then formulated the task as a diagnostic QA prompt: \textit{``Based on the provided clinical data and the Rotterdam criteria, diagnose whether this patient has PCOS. Output YES or NO.''} This adaptation ensures a fair comparison, allowing MedAgents and MDAgents to process the same clinical information as \textsc{Mapis}, differing only in their reasoning architecture.

All experiments ran on a high-performance server equipped with an NVIDIA RTX 5090 GPU (32 GB VRAM) and 64 GB system memory. Machine-learning baselines were implemented in Python 3.9 using scikit-learn and PyTorch 2.0, while LLM-based methods interacted with their official API endpoints.

\subsection{Metrics}
To comprehensively assess diagnostic performance, we employ four standard metrics widely established in PCOS detection literature \cite{tiwari2022sposds}: Accuracy, Precision, Recall, and F1-Score. Accuracy provides a global measure of overall prediction correctness. Recall is of paramount importance in this clinical context to minimize false negatives, ensuring that patients with PCOS are successfully identified for timely intervention. Conversely, Precision evaluates the system's capability to limit false positives, which is essential for mitigating overdiagnosis and preventing inappropriate treatment regimens. Finally, the F1-Score, calculated as the harmonic mean of Precision and Recall, serves as a robust composite indicator that effectively balances the trade-off between diagnostic sensitivity and precision.

\section{Experimental Results}

\begin{table*}[!t]
\centering
\caption{Performance comparison of \textsc{Mapis} versus baselines across public and private datasets. The best and second best results are in \textbf{bold} and \underline{underlined}, respectively.}
\label{tab:performance_comparison}
\adjustbox{width=\textwidth,center}
{%
\begin{tabular}{@{}l|cccc|cccc@{}}
\toprule
\multirow{2}{*}{\textbf{Method}} & 
\multicolumn{4}{c|}{\textbf{Kerala Dataset}} & 
\multicolumn{4}{c}{\textbf{GED Dataset}} \\
\cmidrule(lr){2-5} \cmidrule(lr){6-9}
& \textbf{Acc.} & \textbf{Pre.} & \textbf{Rec.} & \textbf{F1} & 
\textbf{Acc.} & \textbf{Pre.} & \textbf{Rec.} & \textbf{F1} \\
\midrule
% Traditional ML Methods (Only for PCOS dataset)
\multicolumn{9}{@{}l}{\quad\textit{\textbf{Machine Learning Methods}}} \\
Bayesian & 86.24 & 76.92 & \underline{83.33} & 80.00 & 69.41 & 83.32 & 65.71 & 73.05 \\
Random Forest+PCA & 88.07 & \underline{87.10} & 75.00 & 80.60 & 72.94 & 74.25 & 89.74 & 80.87 \\
XGBoost+SCA & \underline{88.99} & 83.33 & \underline{83.33} & \underline{83.33} & 78.20 & 80.50 & 87.00 & 83.60 \\
\midrule
% Single-Agent LLM Methods
\multicolumn{9}{l}{\textit{\textbf{Single-Agent Methods}}} \\
GPT-4o (Zero-shot) & 75.23 & 64.52 & 55.56 & 59.70 & 82.94 & 82.64 & 92.59 & 87.28\\
GPT-4o (CoT) &  69.72 & 71.43 & 13.89 & 23.26 & 80.59 & 80.83 & 90.65 & 85.41 \\
GPT-4o (RAG) & 86.24 & 80.00 & 77.78 & 78.88 & \underline{85.88} & 87.50 & 90.74 & \underline{89.09} \\
GPT-4.1 (Zero-shot)  & 77.98 & 80.00 & 44.44 & 57.16 & 81.76 & 88.12 & 82.41 & 85.25 \\
GPT-4.1 (CoT) & 73.39 & 62.96 & 47.22 & 54.02& 84.71 & 89.42 & 86.11 & 87.73 \\
GPT-4.1 (RAG) & 86.24 & 86.21 & 69.44 & 76.98 & 85.12 & \underline{91.92} & 84.26 & 87.90 \\
Gemini-2.5-flash (Zero-shot) & 66.06 & 48.00 & 33.33 & 39.34 & 70.59 & 81.52 & 69.44 & 74.96 \\
Gemini-2.5-flash (CoT) & 61.47 & 36.36 & 22.22 & 27.59 & 74.71 & 77.31 & 85.19 & 81.05 \\
Gemini-2.5-flash (RAG) & 80.43 & 75.00 & 75.00 & 75.00 & 84.12 & 82.40 & \underline{95.37} & 88.35 \\
Gemini-2.5-pro (Zero-shot) & 80.73 & 74.19 & 63.89 & 68.60 & 81.18 & 85.85 & 84.26 & 85.10 \\
Gemini-2.5-pro (CoT) & 77.78 & 62.22 & 80.00 & 70.00  & 84.71 & 85.34 & 91.67 & 88.44 \\
Gemini-2.5-pro (RAG) & 87.16 & \textbf{92.31} & 66.67 & 77.41 & 85.21 & 84.75 & 93.46 & 88.96 \\
\midrule
% Multi-Agent Methods (Only for GEDD dataset)
\multicolumn{9}{l}{\textit{\textbf{Multi-Agent Methods}}} \\
MDAgent & 82.57 & 86.96 & 55.56 & 67.80 & 82.94 & 83.76 & 90.74 & 87.11 \\
MedAgent & 87.16 & 84.38 & 75.00 & 79.41 & 84.71 & 81.54 & \textbf{98.51} & 89.08 \\
\midrule
% Our Method
\textbf{Ours(MAPIS)}& 
\textbf{89.91} & 
83.78 & 
\textbf{86.11} & 
\textbf{84.93} & 
\textbf{91.76} & 
\textbf{93.52} & 
93.52 & 
\textbf{93.52} \\
\bottomrule
\end{tabular}
}
\end{table*}

\subsection{Main Results}
Table \ref{tab:performance_comparison} presents a comprehensive performance comparison between \textsc{Mapis} and baseline methods across the public Kerala dataset and our private GED. \textsc{Mapis} achieved the best or second-best performance across multiple metrics in both datasets, with only two exceptions. Specifically, on the public benchmark, \textsc{Mapis} establishes a new state-of-the-art, surpassing the runner-up by 1.60\% in F1-score and securing a 2.78\% lead in Recall. Crucially, this advantage is magnified on the private clinical dataset, where \textsc{Mapis} outperforms the nearest competitor by substantial margins: 4.43\% in F1-score and 5.88\% in Accuracy. This consistent superiority validates the efficacy of our guideline-driven multi-agent architecture in navigating diverse clinical scenarios and data distributions.

\subsubsection{Comparison with Machine Learning Methods}\label{vsmachine}
Among traditional machine learning baselines, XGBoost+SCA demonstrates the strongest performance, achieving a peak Accuracy of 88.99\% on the Kerala Dataset. However, its robustness falters on the more complex clinical data, with Accuracy dropping significantly to 78.20\% on the GED Dataset. In stark contrast, \textsc{Mapis} exhibits superior generalization. On the challenging GED dataset, our framework outperforms XGBoost+SCA by a substantial margin of 13.56\% in Accuracy, confirming the necessity of semantic reasoning over pure statistical fitting for clinical diagnostics.

\subsubsection{Comparison with Single-Agent Methods}\label{vssingle}
We evaluate single-agent approaches across two dimensions: prompting strategies and model backbones. 

\paragraph{Impact of Prompting Strategies}
Zero-shot prompting proves inadequate for clinical risk assessment, evidenced by GPT-4o's critically low Recall of 13.89\% on the public dataset. Counter-intuitively, employing CoT prompting frequently degrades performance compared to zero-shot baselines—most notably, GPT-4o's F1-score drops precipitously from 59.7\% to 23.3\% on the public dataset—likely due to ungrounded intermediate hallucinations. Consequently, RAG consistently emerges as the optimal strategy. By anchoring reasoning in external knowledge, RAG boosts the F1-score of GPT-4o to 78.88\%.

\paragraph{Model Capabilities and Ceiling}
Under the RAG configuration, Gemini-2.5-Pro and GPT-4o emerge as the top-performing backbones. Gemini-2.5-pro (RAG) achieves the highest single-agent Accuracy of 87.16\% on the public dataset, demonstrating exceptional reliability in minimizing false positives. Notably, the lightweight Gemini-2.5-flash exhibits the most dramatic benefit from knowledge injection, yielding a 18.96\% improvement in Accuracy. However, even the best-performing single-agent baseline (GPT-4o RAG) hits a discernible performance ceiling. \textsc{Mapis} breaks this bottleneck, outperforming the best single-agent approach by substantial margins of 6.05\% in F1-score on the Kerala dataset and 4.43\% on the GED dataset, quantitatively demonstrating the superiority of our decoupled multi-agent architecture over monolithic contexts.

\subsubsection{Comparison with Multi-Agent Methods}\label{vsmulti}
Our experiments reveal that generalist multi-agent frameworks do not consistently outperform optimized Single-Agent RAG baselines. As benchmarked in Table \ref{tab:performance_comparison}, MDAgent achieves an F1-score of only 67.80\% on the public Kerala dataset, significantly trailing the single-agent GPT-4o (RAG) (78.88\%). Similarly, MedAgent fails to establish a decisive lead, merely converging with the Accuracy of the single-agent Gemini-2.5-pro (RAG) at 87.16\%. This suggests that without domain-specific constraints, the heuristic consensus mechanisms employed by generalist frameworks yield diminishing returns in strictly protocol-driven tasks. \textsc{Mapis} resolves this dichotomy by enforcing a deterministic clinical workflow. By anchoring agent collaboration in the specialized knowledge graph rather than unstructured debate, our system surpasses the best multi-agent baseline (MedAgent) by substantial margins: 5.52\% in F1-score on the public dataset and 4.44\% on the private clinical dataset. 

\subsection{Ablation Study}
We restricted our extensive ablation studies to the private clinical dataset, leveraging its comprehensive 108 feature space to rigorously evaluate each system component. This dataset provides the requisite granularity, specifically complete biochemical panels, imaging parameters, and clinical assessments, essential for validating the Rotterdam-based three-step diagnostic framework. In contrast, the public Kerala dataset lacks critical biomarkers and morphological metrics, rendering it insufficient for validating architectural mechanisms explicitly engineered for guideline adherence. We stratified our analysis along two dimensions: Backbone Adaptability, assessing the robustness of \textsc{Mapis} across diverse LLM architectures; and Component Efficacy, deconstructing specific system contributions to isolate the impact of knowledge injection and workflow orchestration.

\begin{table}[!t]
\centering
\caption{Performance of \textsc{Mapis} with different backbone models}
\label{tab:backbone_comparison}
\begin{tabular}{@{}lcccc@{}}
\toprule
\textbf{Backbone Model} & \textbf{Acc.} & \textbf{Pre.} & \textbf{Rec.} & \textbf{F1} \\
\midrule
\multicolumn{5}{l}{\textit{\textbf{Closed-source Models}}} \\
MA-PCOS (GPT-4.1) & \textbf{91.76} & 93.52 & \textbf{93.52} & \textbf{93.52} \\
MA-PCOS (GPT-4o) & 90.59 & \textbf{94.44} & 91.07 & 92.72 \\
MA-PCOS (Gemini-2.5-pro) & 88.82 & 90.83 & 91.67 & 91.19 \\
MA-PCOS (Gemini-2.5-flash) & 88.24 & 90.00 & 91.67 & 90.83 \\
\bottomrule
\end{tabular}
\end{table}

\subsubsection{Backbone Model Analysis}
Table \ref{tab:backbone_comparison} benchmarks the adaptability of \textsc{Mapis} across diverse backbone LLMs. The results demonstrate that our framework functions as a universal performance amplifier, consistently enhancing diagnostic capabilities regardless of the model's native parameter size. Notably, \textsc{Mapis} (GPT-4.1) emerges as the optimal configuration. While \textsc{Mapis} (GPT-4o) attains a marginally higher Precision (94.44\%), GPT-4.1 secures the highest Accuracy ($91.76\%$) and F1-score (93.52\%), thereby justifying its selection as our primary backbone. Remarkably, even the lightweight Gemini-2.5-flash achieves a robust Accuracy of 88.24\% within our framework, surpassing much larger single-agent models, which further validates that performance superiority stems from our structured multi-agent architecture rather than raw model capacity.

\begin{table}[!t]
\centering
\caption{Detailed ablation metrics evaluating component contributions.}
\label{tab:ablation_study}
\begin{tabular}{@{}lcccc@{}}
\toprule
\textbf{System Configuration} & \textbf{Acc.} & \textbf{Pre.} & \textbf{Rec.} & \textbf{F1} \\
\midrule

% Full System (Baseline)
\textcolor{black}{\textbf{Full System (Ours)}} & 
\textcolor{black}{\textbf{91.76}} & 
\textcolor{black}{\textbf{93.52}} & 
\textcolor{black}{\textbf{93.52}} & 
\textcolor{black}{\textbf{93.52}} \\
\midrule

% Knowledge Graph Ablation
\multicolumn{5}{l}{\textit{\textbf{Knowledge Enhancement Ablation}}} \\
w/o Knowledge Graph & 85.29 & 87.96 & 88.79 & 88.36 \\
w/ Traditional RAG & 87.65 & 90.74 & 89.91 & 90.34 \\
\midrule

% Framework Architecture Ablation  
\multicolumn{5}{l}{\textit{\textbf{Assessment Framework Ablation}}} \\
Single Agent Assessment & 85.29 & 93.52 & 84.87 & 88.99 \\
\midrule

% Exclusion Phase Ablation
\multicolumn{5}{l}{\textit{\textbf{Exclusion Phase Ablation}}} \\
w/o Exclusion Phase & 86.47 & 86.96 & 92.59 & 89.73 \\

\bottomrule
\end{tabular}
\end{table}

\begin{figure*}[H] % Change [t] to [htbp]
    \centering
    \includegraphics[width=0.6\textwidth]{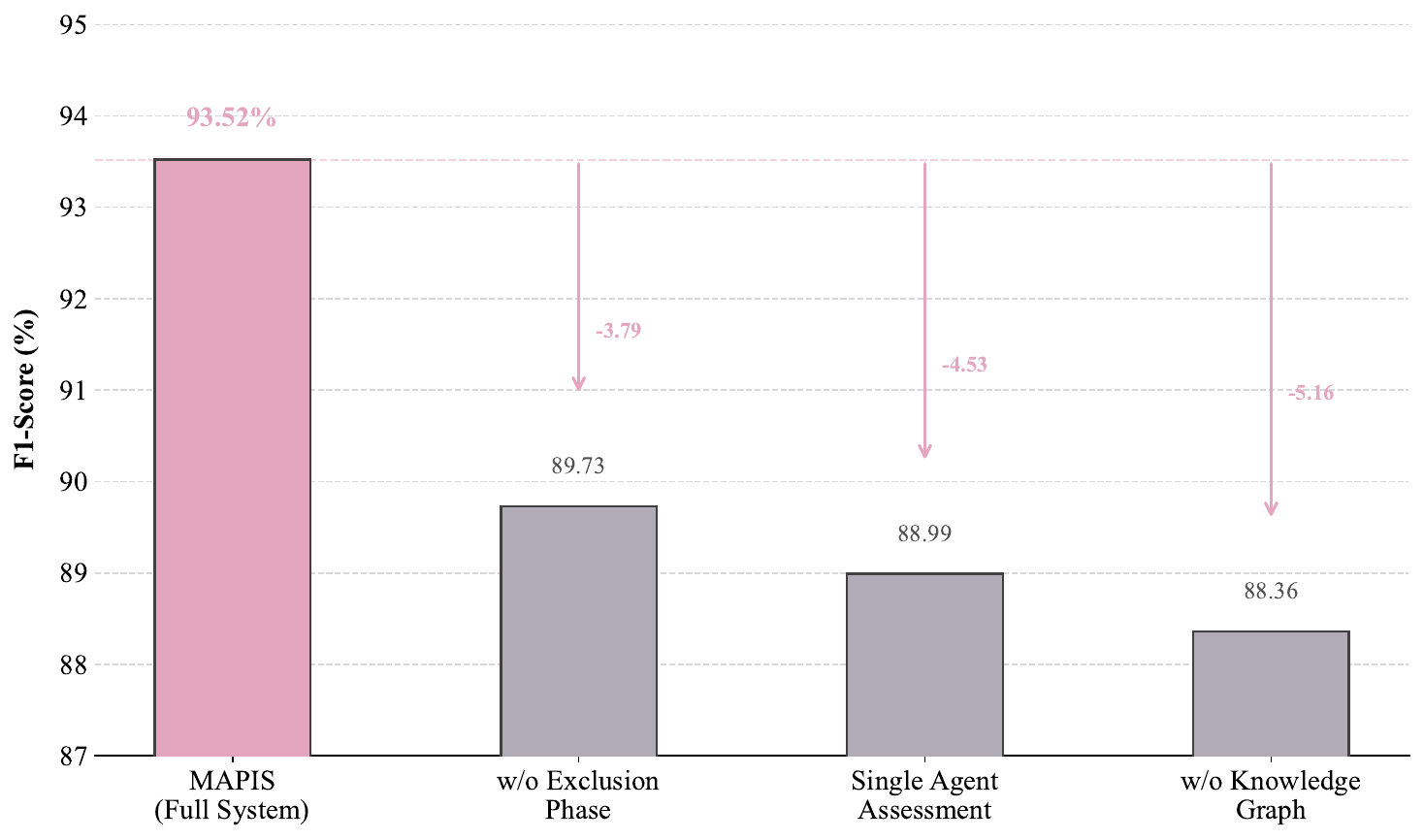}
    \caption{Quantifying the impact of architectural ablation on F1-score.}
    \label{figabla}
\end{figure*}

\subsubsection{System Component Analysis}
To dissect the contribution of specific architectural modules to diagnostic performance, we evaluated four ablation variants against the full \textsc{Mapis} system, as detailed in Table \ref{tab:ablation_study}. Figure \ref{figabla} visually quantifies the impact of removing these key components, illustrating the degradation in F1-score associated with each architectural simplification. The variants include: 1) \textit{w/o Knowledge Graph}, removing the domain KG entirely; 2) \textit{w/ Traditional RAG}, substituting our structured KG with standard semantic retrieval; 3) \textit{Single Agent Assessment}, collapsing the collaborative workflow into a monolithic step; and 4) \textit{w/o Exclusion Phase}, omitting the differential diagnosis protocol.

As illustrated by the steepest drop in Figure \ref{figabla}, the complete removal of the knowledge module precipitates the most severe performance penalty, reducing Accuracy by 6.47\% and the F1-score by 5.16\%. This confirms that domain knowledge grounding is not merely additive but indispensable for reliable medical reasoning. Furthermore, replacing our structured KG with standard RAG still incurs a 4.11\% Accuracy deficit. This finding highlights a critical distinction: while unstructured text retrieval provides context, the structured relationship modeling inherent in our KG is required to navigate complex, interdependent diagnostic criteria efficiently. Collapsing the three-step workflow into a single-agent assessment triggers a distinct failure mode: a drastic 8.65\% collapse in Recall, accompanied by a 4.53\% drop in F1-score. This degradation suggests that without a structured workflow, a monolithic agent succumbs to conservative bias—failing to identify positive cases when faced with ambiguity. Our collaborative architecture prevents this by distributing cognitive load across specialized agents, ensuring that subtle positive evidence is captured rather than overlooked. Finally, omitting the differential diagnosis step, the system experiences a 5.29\% Accuracy reduction and 3.79\% F1 decline. Critically, we observe a distinct failure mode characterized by a sharp decline in Precision of 6.56\%, while Recall remains relatively stable. This sharp decline confirms that the Exclusion Phase functions as a necessary ``safety valve,'' strictly enforcing the clinical mandate to rule out other causes prior to confirmation. 

These results highlight that all strategies in our framework, domain knowledge graph, Guideline-based Multi-Agent Workflow, and Differential Exclusion, play essential and complementary roles in PCOS detection. The KG provides structured medical grounding, the Workflow ensures systematic evidence gathering according to Rotterdam criteria, and the Exclusion phase validates diagnostic conclusions against mimics. The synergy among these strategies significantly enhances the framework’s robustness in clinical diagnosis, as evidenced by the performance degradation when any strategy is disabled.
\begin{figure*}[htbp!]
\centering
    % width=\textwidth 确保图片宽度占满整个页面宽度（两栏总宽）
    \includegraphics[width=\textwidth]{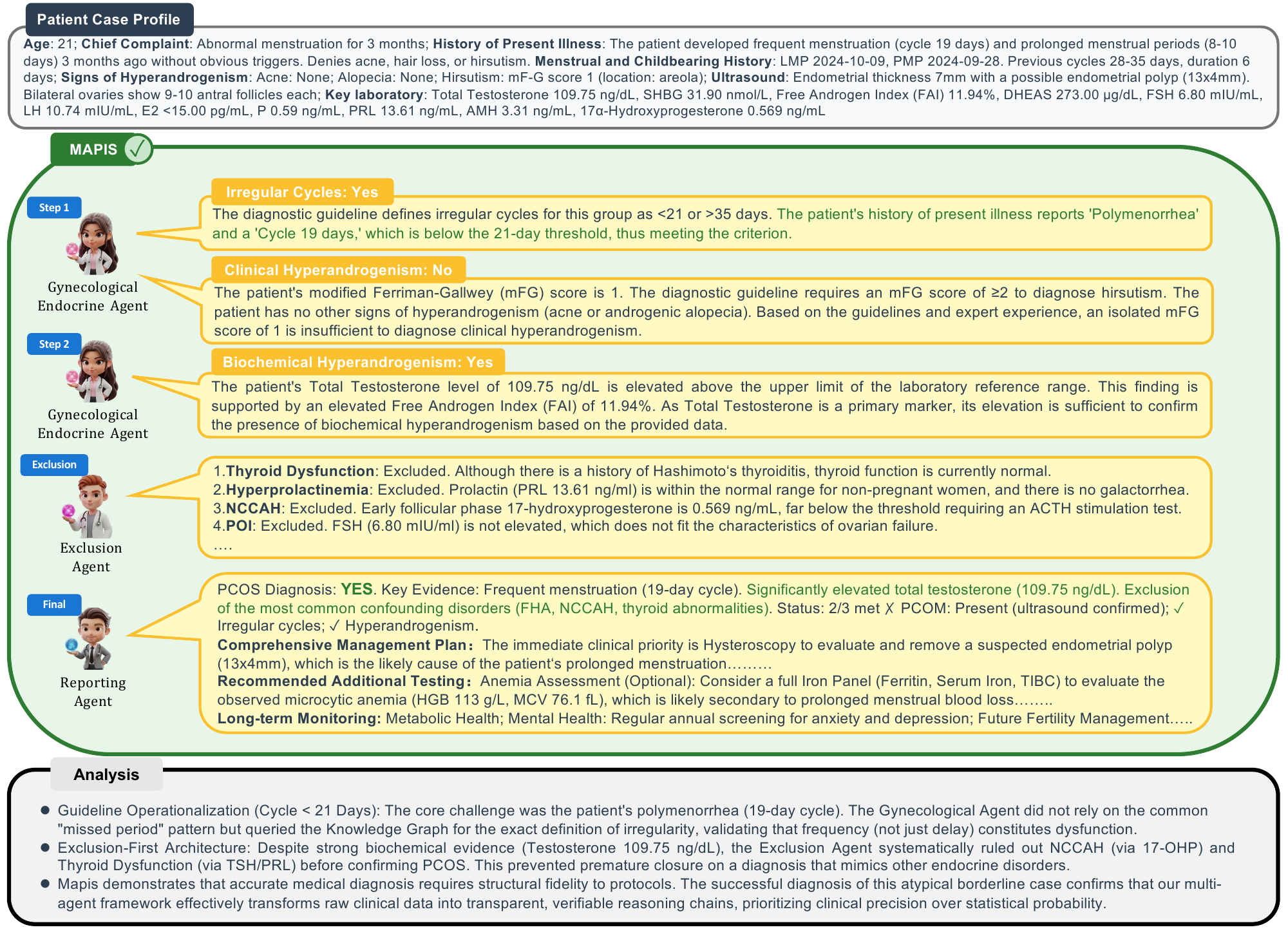}
%\captionsetup{justification=centering}
\caption{Case study 1.}
\label{figcase1}
\end{figure*}

\begin{figure*}[t!]
\centering
    % width=\textwidth 确保图片宽度占满整个页面宽度（两栏总宽）
    \includegraphics[width=\textwidth]{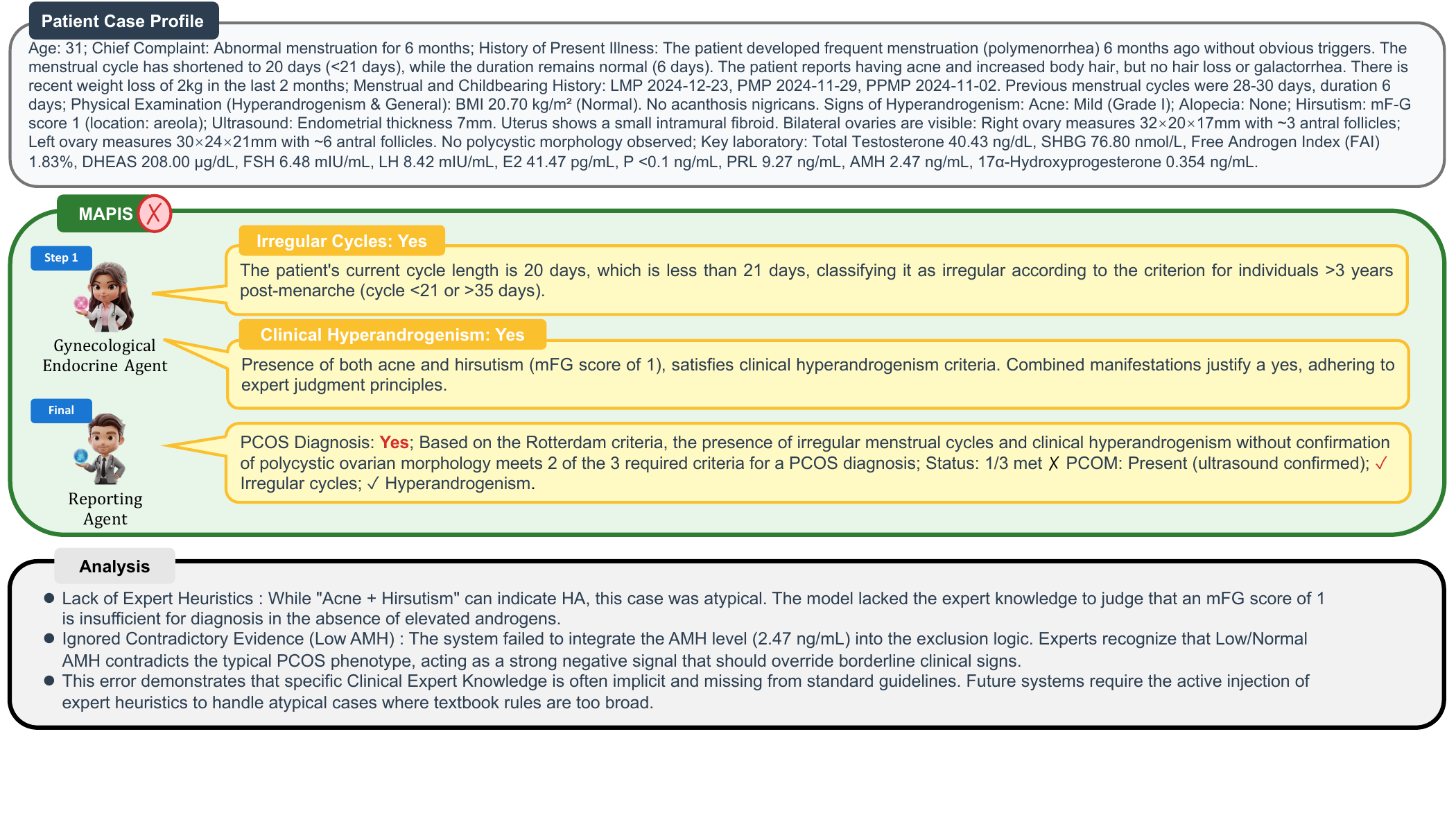}
%\captionsetup{justification=centering}
\caption{Case study 2.}
\label{figcase2}
\end{figure*}
\section{Case Study}
We selected two representative cases from our final results, one successful diagnosis of an atypical phenotype (see Figure \ref{figcase1}) and one false positive (see Figure \ref{figcase2}), to analyze the reasoning capabilities and boundaries of \textsc{Mapis}. Unimportant information has been removed from the texts for a better demonstration.

As shown in Figure \ref{figcase1}, in the successful case 1, the system demonstrated robust guideline operationalization by correctly identifying polymenorrhea as a valid diagnostic criterion, overcoming the common bias in generalist models that associate PCOS primarily with delayed cycles. Our gynecological agent leveraged the knowledge graph to retrieve the precise definition , validating the symptom despite its atypicality. Subsequently, the exclusion agent focused strictly on differential diagnosis, systematically ruling out mimics like NCCAH and thyroid dysfunction to clear the diagnostic path. Finally, the reporting agent aggregated these hierarchical findings and executed the ``two-out-of-three'' logic; despite the absence of polycystic ovarian morphology, it correctly prioritized the combination of irregular cycles and biochemical hyperandrogenism to synthesize the final positive diagnosis.

Conversely, the incorrect Case 2 (Figure \ref{figcase2}) exposes the ``rigidity trap'' inherent in explicit guideline execution. Here, \textsc{Mapis} produced a false positive by mechanically interpreting borderline signs, mild acne and an mF-G score of 1, as definitive clinical hyperandrogenism, failing to exercise the clinical discretion that typically dismisses such weak signals in isolation. The system lacked the implicit expert heuristic to weigh conflicting evidence; specifically, it ignored the strong negative signal from the normal AMH and testosterone levels , which a human expert would use to veto the diagnosis due to the correlation between PCOS and elevated ovarian reserve. This reveals a critical limitation: while knowledge graphs ensure protocol compliance, effectively navigating these borderline cases requires ``tacit knowledge'' (specifically expert experience) to resolve conflicts between qualitative borderline signs and quantitative biochemical negatives.

\section{Discussion}

\subsection{Performance Analysis and Advantages}
The consistent performance advantage of \textsc{Mapis} over diverse baselines stems from a fundamental paradigm shift. We analyze this superiority across three dimensions corresponding to our experimental comparisons:

Traditional machine learning models expose limitations: they heavily rely on extensive labeled data and struggle with the nuanced semantic understanding required for PCOS diagnosis. Diagnostic environments demand a comprehensive integration of extensive data, such as detailed menstrual histories. \textsc{Mapis} overcomes this by leveraging semantic reasoning powered by Large Language Models (LLMs) and grounded in a domain Knowledge Graph. This approach facilitates zero-shot diagnostic inference, effectively shifting the focus from merely fitting numerical patterns to synthesizing comprehensive clinical evidence.

\begin{figure*}[htbp]   
  \centering            
  \includegraphics[width=0.7\textwidth]{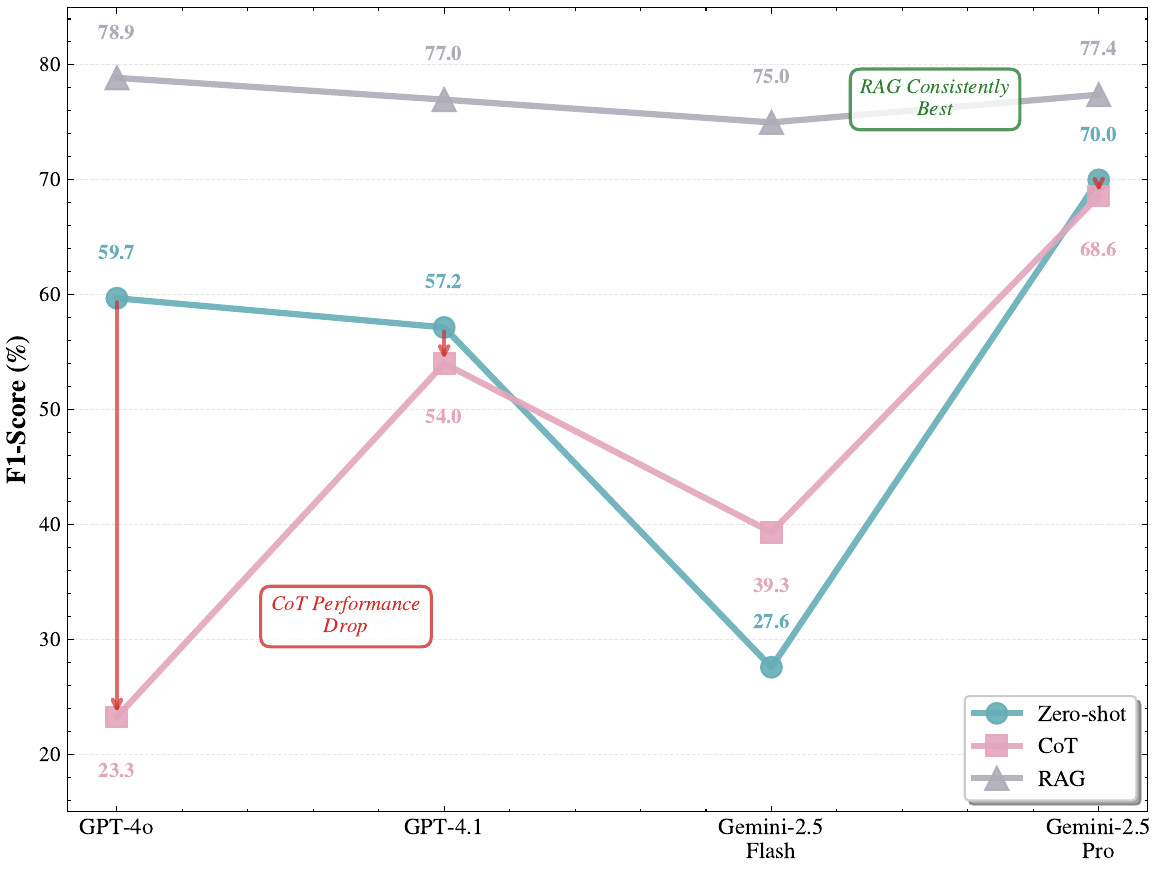}
    \caption{Performance comparison of Single-Agent baselines across different prompting strategies on public Datasets.}
  \label{fig6}         
\end{figure*}

Single-agent often falter due to hallucinations when processing complex PCOS criteria within a single context window, leading to insufficient analysis. As illustrated in Figure \ref{fig6}, we observe a counter-intuitive phenomenon where advanced prompting strategies negatively impact performance. Specifically, GPT-4o's F1-score drops precipitously from 59.7\% to 23.3\% on the public dataset when switching from Zero-shot to CoT. This paradoxical decline aligns with recent findings that ungrounded CoT can induce hallucinations in specialized domains~\cite{tang-etal-2024-medagents,bubeck2023sparks}, where the model fabricates plausible-sounding but medically inaccurate intermediate steps. \textsc{Mapis} demonstrates structural superiority: rather than relying on prompt instructions to implicitly simulate the ``two-out-of-three'' Rotterdam criteria, our framework explicitly operationalizes them into a decoupled pipeline. By assigning distinct cognitive tasks to specialized agents , we ensure that the diagnostic logic is strictly enforced sequentially rather than being statistically approximated.

Multi-agent baselines are primarily designed for general Medical QA rather than specialized PCOS diagnosis. They rely on heuristic consensus mechanisms (e.g., majority voting or debate) optimized for open-ended queries. However, clinical diagnosis requires the deterministic operationalization of protocols, not probabilistic consensus—a voting mechanism cannot replace the mandatory ``exclusion of other causes.'' \textsc{Mapis} resolves this dichotomy by shifting the paradigm from ``unstructured collaboration'' to a ``guideline-operationalized workflow.'' By anchoring reasoning in our specialized knowledge graph, our system surpasses generalist agents by substantial margins, confirming that in specialized medical domains, performance superiority derives not from the mere multiplicity of agents, but from the structural fidelity of the workflow to validated protocols.

\subsection{Limitations} 
Despite the promising diagnostic performance, our study is subject to three primary limitations:

\paragraph{Computational Overhead vs. Clinical Safety} The enhanced interpretability of \textsc{Mapis} incurs a computational premium, averaging 10,913 tokens and 39.8 seconds per case—representing an approximate two-fold increase in cost and a four-fold increase in latency compared to Single-Agent baselines (5,178 tokens, 9.7s). While this latency is clinically negligible compared to the morbidity costs of diagnostic errors, it challenges real-time scalability. Future research must prioritize prompt compression and reasoning distillation to optimize the efficiency-accuracy trade-off for resource-constrained primary care settings.
%The enhanced interpretability and precision of \textsc{Mapis} incur a substantial computational premium. Unlike single-turn inference, our collaborative architecture necessitates extensive inter-agent communication and hierarchical context injection. Benchmarking reveals that \textsc{Mapis} consumes an average of 10,913 tokens with a latency of 39.8 seconds per case—an approximate $2\times$ increase in cost and $4\times$ increase in time compared to the Single-Agent RAG baseline (5,178 tokens, 9.7 seconds). While this latency precludes real-time applications, it remains clinically negligible when weighed against the profound financial and morbidity costs of diagnostic errors, or the typical duration of specialist consultations. We thus frame this overhead not as an inefficiency, but as a necessary ``Cost of Quality'' for high-stakes decision-making. Nonetheless, future research must prioritize prompt compression and reasoning distillation to optimize the efficiency-Accuracy trade-off for deployment in resource-constrained primary care settings.
\paragraph{Dependency on Proprietary APIs} Our reliance on closed-source models ensures state-of-the-art reasoning but introduces constraints regarding data privacy and cost scalability. We have not yet evaluated \textsc{Mapis} on locally deployed open-source models (e.g., Llama-3), which are preferred in hospital environments to ensure strict data sovereignty. Verifying the framework's performance on lower-parameter, ``On-Premise'' models remains a critical direction for secure clinical deployment.
%Our evaluation relies exclusively on high-performance proprietary models accessed via APIs. While these models provide state-of-the-art reasoning capabilities, their closed nature introduces dependencies regarding data privacy, API stability, and cost scalability. We did not evaluate locally deployed open-source models (e.g., Llama-3 or Mistral), which are often preferred in hospital environments to ensure strict compliance with data sovereignty regulations. Consequently, the performance of \textsc{Mapis} on lower-parameter, offline models remains unverified, marking a critical direction for future ``On-Premise AI'' research. 
\paragraph{Single-Center Retrospective Bias} Although validated by senior specialists, the study is retrospective and restricted to a tertiary center in Shenzhen. This design introduces potential selection bias and limits generalizability regarding phenotypic heterogeneity, such as the well-documented ethnic variations in BMI and hirsutism thresholds. Establishing external validity across diverse global demographics requires future large-scale, multi-center prospective validation.
%Although diagnostic Accuracy was validated by senior gynecological endocrinologists, the study design remains retrospective and single-centered. Data collection was restricted to a tertiary care center in Shenzhen, potentially introducing selection bias and limiting demographic diversity. Specifically, the lack of multi-site validation may overlook phenotypic heterogeneity, such as the well-documented variations in BMI and hirsutism thresholds between Asian and Western populations. Therefore, while the internal validity of our framework is established, confirming its external validity across diverse global demographics requires large-scale, multi-center prospective cohorts.

\section{Conclusion}
In this paper, we introduced \textsc{Mapis}, the first multi-agent framework explicitly engineered for guideline-based PCOS diagnosis. Grounded in the International Evidence-based Guideline 2023, our architecture orchestrates structured agent collaboration to execute the three-step Rotterdam evaluation sequence and mandatory differential exclusion workflows, all supported by a hierarchical PCOS-specific knowledge graph. Comprehensive benchmarking on both public and specialized clinical datasets demonstrates \textsc{Mapis}'s substantial superiority over traditional machine learning, single-agent, and representative medical multi-agent systems. Furthermore, ablation studies confirm that the architectural decoupling of cognitive tasks contributes essential, synergistic value to the overall diagnostic precision. Through zero-shot inference enabled by knowledge operationalization, \textsc{Mapis} produces transparent reasoning chains verifiable by clinicians, ensuring systematic adherence to international diagnostic standards. These capabilities underscore a critical thesis: bridging the gap between computational intelligence and clinical deployment requires a paradigm shift from generic AI models to disease-specific, workflow-faithful architectures. Ultimately, this work advances the frontier of automated clinical reasoning by maintaining the interpretability and guideline compliance critical for patient safety, thereby opening new avenues for reliable LLM-assisted diagnosis in complex endocrine disorders.

\bibliographystyle{cas-model2-names}

\bibliography{cas-refs}

@misc{teede2023international,
  title={International evidence-based guideline for the assessment and management of polycystic ovary syndrome 2023},
  author={Teede, Helena and Tay, Chau Thien and Laven, Joop and Dokras, Anuja and Moran, Lisa and Piltonen, Terhi and Costello, Michael and Boivin, Jacky and Redman, Leanne and Boyle, Jacqueline and Norman, Robert and Mousa, Aya and Joham, Anju},
  year={2023},
  note = {https://www.rcog.org.uk/guidance/browse-all-guidance/other-guidelines-and-reports/international-evidence-based-guideline-on-polycystic-ovary-syndrome/, 2023, (accessed 25 December 2024)}
}

@inproceedings{wu-etal-2025-medical,
    title = "Medical Graph {RAG}: Evidence-based Medical Large Language Model via Graph Retrieval-Augmented Generation",
    author = "Wu, Junde  and
      Zhu, Jiayuan  and
      Qi, Yunli  and
      Chen, Jingkun  and
      Xu, Min  and
      Menolascina, Filippo  and
      Jin, Yueming  and
      Grau, Vicente",
    booktitle = "Proceedings of the 63rd Annual Meeting of the Association for Computational Linguistics (Volume 1: Long Papers)",
    month = jul,
    year = "2025",
    address = "Vienna, Austria",
    publisher = "Association for Computational Linguistics",
    url = "https://aclanthology.org/2025.acl-long.1381/",
    doi = "10.18653/v1/2025.acl-long.1381",
    pages = "28443--28467",
    ISBN = "979-8-89176-251-0",
}

@misc{pcosdataset,
    title = {Polycystic ovary syndrome (PCOS) [dataset]},
    author = {Prasoon, Kottarathil},
    year = {2020},
    note = {Kaggle, https://www.kaggle.com/datasets/prasoonkottarathil/polycystic-ovary-syndrome-pcos},
}

@article{18subha2024computational,
title = {Computational intelligence for early detection of infertility in women},
journal = {Engineering Applications of Artificial Intelligence},
volume = {127},
pages = {107400},
year = {2024},
issn = {0952-1976},
doi = {https://doi.org/10.1016/j.engappai.2023.107400},
author = {Subha R. and Nayana B.R. and Rekha Radhakrishnan and Sumalatha P.},
}

@article{84kumari2023sms,
author = {Kumari, Ritika and Singh, Jaspreeti and Gosain, Anjana},
title = {SmS: SMOTE-stacked hybrid model for diagnosis of polycystic ovary syndrome using feature selection method},
year = {2023},
issue_date = {Sep 2023},
publisher = {Pergamon Press, Inc.},
address = {USA},
volume = {225},
number = {C},
issn = {0957-4174},
doi = {10.1016/j.eswa.2023.120102},
journal = {Expert Syst. Appl.},
month = sep,
numpages = {10}
}

@article{34elmannai2023polycystic,
  title={Polycystic ovary syndrome detection machine learning model based on optimized feature selection and explainable artificial intelligence},
  author={Elmannai, Hela and El-Rashidy, Nora and Mashal, Ibrahim and Alohali, Manal Abdullah and Farag, Sara and El-Sappagh, Shaker and Saleh, Hager},
  journal={Diagnostics},
  volume={13},
  number={8},
  pages={1506},
  year={2023},
  publisher={MDPI},
doi={https://doi.org/10.3390/diagnostics13081506}
}

@ARTICLE{57nasim2022novel,
  author={Nasim, Shazia and Almutairi, Mubarak Saad and Munir, Kashif and Raza, Ali and Younas, Faizan},
  journal={IEEE Access}, 
  title={A Novel Approach for Polycystic Ovary Syndrome Prediction Using Machine Learning in Bioinformatics}, 
  year={2022},
  volume={10},
  number={},
  pages={97610-97624},
  doi={10.1109/ACCESS.2022.3205587}}

@article{li2025intelligent,
  title={Intelligent detection for Polycystic Ovary Syndrome (PCOS): Taxonomy, datasets and detection tools},
  author={Li, Meng and He, Zanxiang and Shi, Liyun and Lin, Mengyuan and Li, Minge and Cheng, Yanjun and Liu, Hongwei and Xue, Lei and Said, Kabir Sulaiman and Yusuf, Murtala and others},
  journal={Computational and Structural Biotechnology Journal},
  year={2025},
  publisher={Elsevier}
}

@INPROCEEDINGS{i-HOPE,
  author={Denny, Amsy and Raj, Anita and Ashok, Ashi and Ram, C Maneesh and George, Remya},
  booktitle={TENCON 2019 - 2019 IEEE Region 10 Conference (TENCON)}, 
  title={i-HOPE: Detection And Prediction System For Polycystic Ovary Syndrome (PCOS) Using Machine Learning Techniques}, 
  year={2019},
  volume={},
  number={},
  pages={673-678},
  doi={10.1109/TENCON.2019.8929674}
}

@article{Sinecosine,
  title={Sine cosine algorithm-based feature selection for improved machine learning models in polycystic ovary syndrome diagnosis},
  author={Rajput, Ishwari Singh and Tyagi, Sonam and Gupta, Aditya and Jain, Vibha},
  journal={Multimedia Tools and Applications},
  pages={1--25},
  year={2024},
  publisher={Springer},
  doi = {https://doi.org/10.1007/s11042-024-18213-z}
}

@article{ACritical,
  title={A Critical Study of Polycystic Ovarian Syndrome (PCOS) Classification Techniques},
  author={B. Vikas and B. S. Anuhya and Manaswini Chilla and Sipra Sarangi},
  volume={21},
  pages = {1-7},
  year={2018},
  journal = {IJCEM International Journal of Computational Engineering \& Management},
}

@article{hurst2024gpt,
  title={Gpt-4o system card},
  author={Hurst, Aaron and Lerer, Adam and Goucher, Adam P and Perelman, Adam and Ramesh, Aditya and Clark, Aidan and Ostrow, AJ and Welihinda, Akila and Hayes, Alan and Radford, Alec and others},
  journal={arXiv preprint arXiv:2410.21276},
  year={2024}
}

@article{comanici2025gemini,
  title={Gemini 2.5: Pushing the frontier with advanced reasoning, multimodality, long context, and next generation agentic capabilities},
  author={Comanici, Gheorghe and Bieber, Eric and Schaekermann, Mike and Pasupat, Ice and Sachdeva, Noveen and Dhillon, Inderjit and Blistein, Marcel and Ram, Ori and Zhang, Dan and Rosen, Evan and others},
  journal={arXiv preprint arXiv:2507.06261},
  year={2025}
}

@inproceedings{tang-etal-2024-medagents,
    title = "{M}ed{A}gents: Large Language Models as Collaborators for Zero-shot Medical Reasoning",
    author = "Tang, Xiangru  and
      Zou, Anni  and
      Zhang, Zhuosheng  and
      Li, Ziming  and
      Zhao, Yilun  and
      Zhang, Xingyao  and
      Cohan, Arman  and
      Gerstein, Mark",
    editor = "Ku, Lun-Wei  and
      Martins, Andre  and
      Srikumar, Vivek",
    booktitle = "Findings of the Association for Computational Linguistics: ACL 2024",
    month = aug,
    year = "2024",
    address = "Bangkok, Thailand",
    publisher = "Association for Computational Linguistics",
    url = "https://aclanthology.org/2024.findings-acl.33/",
    doi = "10.18653/v1/2024.findings-acl.33",
    pages = "599--621",
}

@article{kim2024mdagents,
  title={Mdagents: An adaptive collaboration of llms for medical decision-making},
  author={Kim, Yubin and Park, Chanwoo and Jeong, Hyewon and Chan, Yik S and Xu, Xuhai and McDuff, Daniel and Lee, Hyeonhoon and Ghassemi, Marzyeh and Breazeal, Cynthia and Park, Hae W},
  journal={Advances in Neural Information Processing Systems},
  volume={37},
  pages={79410--79452},
  year={2024}
}

@article{aggarwal2023early,
title = {Early identification of PCOS with commonly known diseases: Obesity, diabetes, high blood pressure and heart disease using machine learning techniques},
journal = {Expert Systems with Applications},
volume = {217},
pages = {119532},
year = {2023},
issn = {0957-4174},
doi = {https://doi.org/10.1016/j.eswa.2023.119532},
author = {Shivani Aggarwal and Kavita Pandey},
}

@article{tiwari2022sposds,
title = {SPOSDS: A smart Polycystic Ovary Syndrome diagnostic system using machine learning},
journal = {Expert Systems with Applications},
volume = {203},
pages = {117592},
year = {2022},
issn = {0957-4174},
doi = {https://doi.org/10.1016/j.eswa.2022.117592},
author = {Shamik Tiwari and Lalit Kane and Deepika Koundal and Anurag Jain and Adi Alhudhaif and Kemal Polat and Atef Zaguia and Fayadh Alenezi and Sara A. Althubiti},
}

@article{bubeck2023sparks,
  title={Sparks of artificial general intelligence: early experiments with GPT-4 (2023)},
  author={Bubeck, S{\'e}bastien and Chandrasekaran, Varun and Eldan, Ronen and Gehrke, Johannes and Horvitz, Eric and Kamar, Ece and Lee, Peter and Lee, Yin Tat and Li, Yuanzhi and Lundberg, Scott and others},
  journal={arXiv preprint arXiv:2303.12712},
  volume={1},
  year={2023}
}

@article{teede2018recommendations,
  title={Recommendations from the international evidence-based guideline for the assessment and management of polycystic ovary syndrome},
  author={Teede, Helena J and Misso, Marie L and Costello, Michael F and Dokras, Anuja and Laven, Joop and Moran, Lisa and Piltonen, Terhi and Norman, Robert J},
  journal={European journal of endocrinology},
  volume={189},
  number={2},
  pages={G43–G64},
  year={2023},
  publisher={Oxford University Press},
doi={https://doi.org/10.1093/ejendo/lvad096}
}

@article{azziz2016polycystic,
  title={Polycystic ovary syndrome},
  author={Azziz, Ricardo and Carmina, Enrico and Chen, ZiJiang and Dunaif, Andrea and Laven, Joop SE and Legro, Richard S and Lizneva, Daria and Natterson-Horowtiz, Barbara and Teede, Helena J and Yildiz, Bulent O},
  journal={Nature reviews Disease primers},
  volume={2},
  number={1},
  pages={1--18},
  year={2016},
  publisher={Nature Publishing Group},
doi = {https://doi.org/10.1038/nrdp.2016.57}
}

@article{march2010prevalence,
  title={The prevalence of polycystic ovary syndrome in a community sample assessed under contrasting diagnostic criteria},
  author={March, Wendy A and Moore, Vivienne M and Willson, Kristyn J and Phillips, David IW and Norman, Robert J and Davies, Michael J},
  journal={Human reproduction},
  volume={25},
  number={2},
  pages={544--551},
  year={2010},
  publisher={Oxford University Press}
}

@article{teede2010polycystic,
  title={Polycystic ovary syndrome: a complex condition with psychological, reproductive and metabolic manifestations that impacts on health across the lifespan},
  author={Teede, Helena and Deeks, Amanda and Moran, Lisa},
  journal={BMC medicine},
  volume={8},
  number={1},
  pages={41},
  year={2010},
  publisher={Springer}
}

@article{boomsma2006meta,
  title={A meta-analysis of pregnancy outcomes in women with polycystic ovary syndrome},
  author={Boomsma, CM and Eijkemans, MJC and Hughes, EG and Visser, GHA and Fauser, BCJM and Macklon, NS},
  journal={Human reproduction update},
  volume={12},
  number={6},
  pages={673--683},
  year={2006},
  publisher={Oxford University Press}
}

@article{apridonidze2005prevalence,
  title={Prevalence and characteristics of the metabolic syndrome in women with polycystic ovary syndrome},
  author={Apridonidze, Teimuraz and Essah, Paulina A and Iuorno, Maria J and Nestler, John E},
  journal={The Journal of Clinical Endocrinology \& Metabolism},
  volume={90},
  number={4},
  pages={1929--1935},
  year={2005},
  publisher={Oxford University Press}
}

@article{eshre2004revised,
title = {Revised 2003 consensus on diagnostic criteria and long-term health risks related to polycystic ovary syndrome},
journal = {Fertility and Sterility},
author = {Rotterdam ESHRE/ASRM-Sponsored PCOS Consensus Workshop Group},
volume = {81},
number = {1},
pages = {19-25},
year = {2004},
issn = {0015-0282},
doi = {https://doi.org/10.1016/j.fertnstert.2003.10.004},
}

@article{saini2016gaps,
  title={Gaps in knowledge in diagnosis and management of polycystic ovary syndrome.},
  author={Saini, Shailly and Gibson-Helm, Melanie and Cooney, Laura and Teede, Helena J and Dokras, Anuja},
  journal={Fertility and Sterility},
  volume={106},
  number={3 (Suppl.)},
  pages={e100},
  year={2016},
  publisher={Elsevier}
}

@article{gibson2017delayed,
  title={Delayed diagnosis and a lack of information associated with dissatisfaction in women with polycystic ovary syndrome},
  author={Gibson-Helm, Melanie and Teede, Helena and Dunaif, Andrea and Dokras, Anuja},
  journal={The Journal of Clinical Endocrinology \& Metabolism},
  volume={102},
  number={2},
  pages={604--612},
  year={2017},
  publisher={Endocrine Society Washington, DC}
}

@INPROCEEDINGS{72deshpande2014automated,
  author={Deshpande, Sharvari S. and Wakankar, Asmita},
  booktitle={2014 IEEE International Conference on Advanced Communications, Control and Computing Technologies}, 
  title={Automated detection of Polycystic Ovarian Syndrome using follicle recognition}, 
  year={2014},
  volume={},
  number={},
  pages={1341-1346},
  doi={10.1109/ICACCCT.2014.7019318}}

@article{67zhang2021raman,
title = {Raman spectroscopy of follicular fluid and plasma with machine-learning algorithms for polycystic ovary syndrome screening},
journal = {Molecular and Cellular Endocrinology},
volume = {523},
pages = {111139},
year = {2021},
issn = {0303-7207},
doi = {https://doi.org/10.1016/j.mce.2020.111139},
author = {Xinyi Zhang and Bo Liang and Jun Zhang and Xinyao Hao and Xiaoyan Xu and Hsun-Ming Chang and Peter C.K. Leung and Jichun Tan},
}

@INPROCEEDINGS{65chauhan2021comparative,
  author={Chauhan, Preeti and Patil, Pooja and Rane, Neha and Raundale, Pooja and Kanakia, Harshil},
  booktitle={2021 International Conference on Communication information and Computing Technology (ICCICT)}, 
  title={Comparative Analysis of Machine Learning Algorithms for Prediction of PCOS}, 
  year={2021},
  volume={},
  number={},
  pages={1-7},
  doi={10.1109/ICCICT50803.2021.9510128}}

@INPROCEEDINGS{srivastav2024transfer,
  author={Srivastav, Somya and Guleria, Kalpna and Sharma, Shagun},
  booktitle={2024 2nd International Conference on Intelligent Data Communication Technologies and Internet of Things (IDCIoT)}, 
  title={A Transfer Learning-Based Fine Tuned VGG16 Model for PCOS Classification}, 
  year={2024},
  volume={},
  number={},
  pages={1074-1079},
  doi={10.1109/IDCIoT59759.2024.10467747}
}

@ARTICLE{56lv2022deep,
AUTHOR={Lv, Wenqi  and Song, Ying  and Fu, Rongxin  and Lin, Xue  and Su, Ya  and Jin, Xiangyu  and Yang, Han  and Shan, Xiaohui  and Du, Wenli  and Huang, Qin  and Zhong, Hao  and Jiang, Kai  and Zhang, Zhi  and Wang, Lina  and Huang, Guoliang },
TITLE={Deep Learning Algorithm for Automated Detection of Polycystic Ovary Syndrome Using Scleral Images},
JOURNAL={Frontiers in Endocrinology},
VOLUME={12},
YEAR={2022},
DOI={10.3389/fendo.2021.789878},
ISSN={1664-2392},
}

@article{46abouhawwash2023automatic,
author = {Abouhawwash, Mohamed and Sridevi, S. and Sundararajan, Suma and Pachlor, Rohit and Karim, Faten and Khafaga, Doaa},
year = {2023},
month = {01},
pages = {239-253},
title = {Automatic Diagnosis of Polycystic Ovarian Syndrome Using Wrapper Methodology with Deep Learning Techniques},
volume = {47},
journal = {Computer Systems Science and Engineering},
doi = {10.32604/csse.2023.037812}
}

@article{55zigarelli2022machine,
  title={Machine-aided self-diagnostic prediction models for polycystic ovary syndrome: observational study},
  author={Zigarelli, Angela and Jia, Ziyang and Lee, Hyunsun},
  journal={JMIR Formative Research},
  volume={6},
  number={3},
  pages={e29967},
  year={2022},
  publisher={JMIR Publications Toronto, Canada},
 doi={https://doi.org/10.2196/29967}
}

@article{li2024agent,
  title={Agent hospital: A simulacrum of hospital with evolvable medical agents},
  author={Li, Junkai and Lai, Yunghwei and Li, Weitao and Ren, Jingyi and Zhang, Meng and Kang, Xinhui and Wang, Siyu and Li, Peng and Zhang, Ya-Qin and Ma, Weizhi and others},
  journal={arXiv preprint arXiv:2405.02957},
  year={2024}
}

@article{wang2024beyond,
  title={Beyond direct diagnosis: LLM-based multi-specialist agent consultation for automatic diagnosis},
  author={Wang, Haochun and Zhao, Sendong and Qiang, Zewen and Xi, Nuwa and Qin, Bing and Liu, Ting},
  journal={arXiv preprint arXiv:2401.16107},
  year={2024}
}

@article{lewis2020retrieval,
  title={Retrieval-augmented generation for knowledge-intensive nlp tasks},
  author={Lewis, Patrick and Perez, Ethan and Piktus, Aleksandra and Petroni, Fabio and Karpukhin, Vladimir and Goyal, Naman and K{\"u}ttler, Heinrich and Lewis, Mike and Yih, Wen-tau and Rockt{\"a}schel, Tim and others},
  journal={Advances in neural information processing systems},
  volume={33},
  pages={9459--9474},
  year={2020}
}

@inproceedings{guu2020retrieval,
  title={Retrieval augmented language model pre-training},
  author={Guu, Kelvin and Lee, Kenton and Tung, Zora and Pasupat, Panupong and Chang, Mingwei},
  booktitle={International conference on machine learning},
  pages={3929--3938},
  year={2020},
  organization={PMLR}
}

@article{miao2024integrating,
  title={Integrating retrieval-augmented generation with large language models in nephrology: advancing practical applications},
  author={Miao, Jing and Thongprayoon, Charat and Suppadungsuk, Supawadee and Garcia Valencia, Oscar A and Cheungpasitporn, Wisit},
  journal={Medicina},
  volume={60},
  number={3},
  pages={445},
  year={2024},
  publisher={MDPI}
}

@article{izacard2021unsupervised,
  title={Unsupervised dense information retrieval with contrastive learning},
  author={Izacard, Gautier and Caron, Mathilde and Hosseini, Lucas and Riedel, Sebastian and Bojanowski, Piotr and Joulin, Armand and Grave, Edouard},
  journal={arXiv preprint arXiv:2112.09118},
  year={2021}
}

@INPROCEEDINGS{8530028,
  author={Li, Hongwei and Li, Sirui and Sun, Jiamou and Xing, Zhenchang and Peng, Xin and Liu, Mingwei and Zhao, Xuejiao},
  booktitle={2018 IEEE International Conference on Software Maintenance and Evolution (ICSME)}, 
  title={Improving API Caveats Accessibility by Mining API Caveats Knowledge Graph}, 
  year={2018},
  volume={},
  number={},
  pages={183-193},
  doi={10.1109/ICSME.2018.00028}}

@article{edge2024local,
  title={From local to global: A graph rag approach to query-focused summarization},
  author={Edge, Darren and Trinh, Ha and Cheng, Newman and Bradley, Joshua and Chao, Alex and Mody, Apurva and Truitt, Steven and Metropolitansky, Dasha and Ness, Robert Osazuwa and Larson, Jonathan},
  journal={arXiv preprint arXiv:2404.16130},
  year={2024}
}

% % Here goes the biography details.

% % Here goes the biography details.

\end{document}